\documentclass[11pt]{amsart}
\usepackage{bm,amsthm,amsmath,amssymb,amsfonts,delarray,array}
\usepackage{graphicx}
\usepackage[round]{natbib}
\usepackage[subrefformat=parens]{subcaption}
\usepackage{enumerate}
\usepackage[ruled]{algorithm2e}
\usepackage{fullpage}
\usepackage{comment}
\usepackage{here}
\newcommand{\argmax}{\mathop{\mathrm{argmax}}}
\newcommand{\argmin}{\mathop{\mathrm{argmin}}}
\newcommand{\res}{\mathrm{res}}
\renewcommand{\tilde}{\widetilde}

\newtheorem{proposition}{Proposition}[section]

\theoremstyle{definition}

\newtheorem{remark}{Remark}[section]

\title[]{Learning partially ranked data \\ based on graph regularization}
\author{Kento Nakamura, Keisuke Yano, \and Fumiyasu Komaki}
\address{Department of Mathematical Informatics,\\ Graduate School of Information Science and Technology,\\ The University of Tokyo}
\email{\{kento\_nakamura,yano,komaki\}@mist.i.u-tokyo.ac.jp}
\date{\today.}
\begin{document}

\begin{abstract}
Ranked data appear in many different applications, including voting and consumer surveys.
There often exhibits a situation in which data are partially ranked.
Partially ranked data is thought of as missing data.
This paper addresses parameter estimation for partially ranked data under a (possibly) non-ignorable missing mechanism.
We propose estimators for both complete rankings and missing mechanisms together with a simple estimation procedure.
Our estimation procedure leverages a graph regularization in conjunction with the Expectation-Maximization algorithm.
Our estimation procedure is theoretically guaranteed to have the convergence properties.
We reduce a modeling bias by allowing a non-ignorable missing mechanism.
In addition, we avoid the inherent complexity within a non-ignorable missing mechanism by introducing a graph regularization.
The experimental results demonstrate that the proposed estimators work well under  non-ignorable missing mechanisms.
\end{abstract}

\maketitle

\section{Introduction}
Data commonly come in the form of ranking in preference survey such as voting and consumer surveys.
Asking people to rearrange items according to their preference,
we obtain the collection of rankings.
Several methods for ranked data have been proposed.
\cite{mallows1957non} proposed a parametric model, now called the Mallows model;
\cite{diaconis1989generalization}
developed a spectral analysis for ranked data;
Recently, the analysis of ranked data has gathered much attention in machine learning community (see \cite{liu2011learning,furnkranz2011preference}).
See Section \ref{section: literature}
for more details.

Partially ranked data is often observed
in real data analysis.
This is because one does not necessarily express his or her preference completely;
for example,
according to the election records of
the American Psychological Association collected in 1980,
one-third of ballots provided full preferences for five candidates,
and the rest provided only top-$t$ preferences with $t=1,2,3$
(see Section 2A in \cite{diaconis1989generalization});
Data are commonly of partially ranked in movie ratings
because respondents usually know only a few movie titles
among a vast number of movies.
Therefore, 
analyzing partially ranked data efficiently
extends the range of application
of statistical methods for ranked data.

Partially ranked data is thought of as missing data.
We can naturally consider that there exists a latent complete ranking behind a partial ranking as discussed in \cite{lebanon2008non}.
The existing studies for partially ranked data make the Missing-At-Random (MAR) assumption,
that is,
an assumption that the missing mechanism generating partially ranked data is ignorable;
Under the MAR assumption,
\cite{busse2007cluster} and \cite{meilua2010dirichlet} 
leverage an extended distance for partially ranked data;
\cite{lu2011learning} introduces a probability model for partially ranked data. 
However, an improper application of the MAR assumption may lead to a relatively large estimation error as argued in the literature on missing data analysis (\cite{little2014statistical}).
In the statistical sense,
if the missing mechanism is non-ignorable,
using the MAR assumption 
is equivalent to
using a misspecified likelihood function, which causes significantly biased parameter estimation and prediction.
In fact, \cite{marlin2009collaborative} 
points out that there occurs a violation of the MAR assumption in music rankings.

This paper addresses learning the distribution of complete and partial rankings based on partially ranked data under a (possibly) non-ignorable missing mechanism.
Our approach includes 
estimating a missing mechanism.
However, estimating a missing mechanism has an intrinsic difficulty.
Consider a top-$t$ ranking of $r$ items.
Length $t$ characterizes
the missing pattern generating a top-$t$ ranking from a complete ranking with $r$ items.
It requires $r!(r-2)$ parameters 
to fully parameterize the missing mechanism 
since $r!$ multinomial distributions with $r-1$ categories
models the missing mechanism.
Note that the number of complete rankings is $r!$.
A large number of parameters cause over-fitting 
especially when the sample size is small.
To avoid over-fitting, 
we introduce an estimation method 
leveraging the recent graph regularization technique (\cite{hallac2015network}) together with the Expectation-Maximization (EM) algorithm.
The numerical experiments using simulation data as well as applications to real data 
indicate that our proposed estimation method works well especially under non-ignorable missing mechanisms.

\subsection{Contribution}

In this paper, we propose estimators for the distribution of a latent complete ranking
and for a missing mechanism.
To this end,
we employ both a latent variable model and 
a recently developed graph regularization.
Our proposal has two merits:
First, we allow a missing mechanism to be non-ignorable by fully parameterizing it.
Second, we reduce over-fitting due to the complexity of missing mechanisms by exploiting a graph regularization method.

\begin{figure}[h]
    \centering
    \includegraphics[width=7cm]{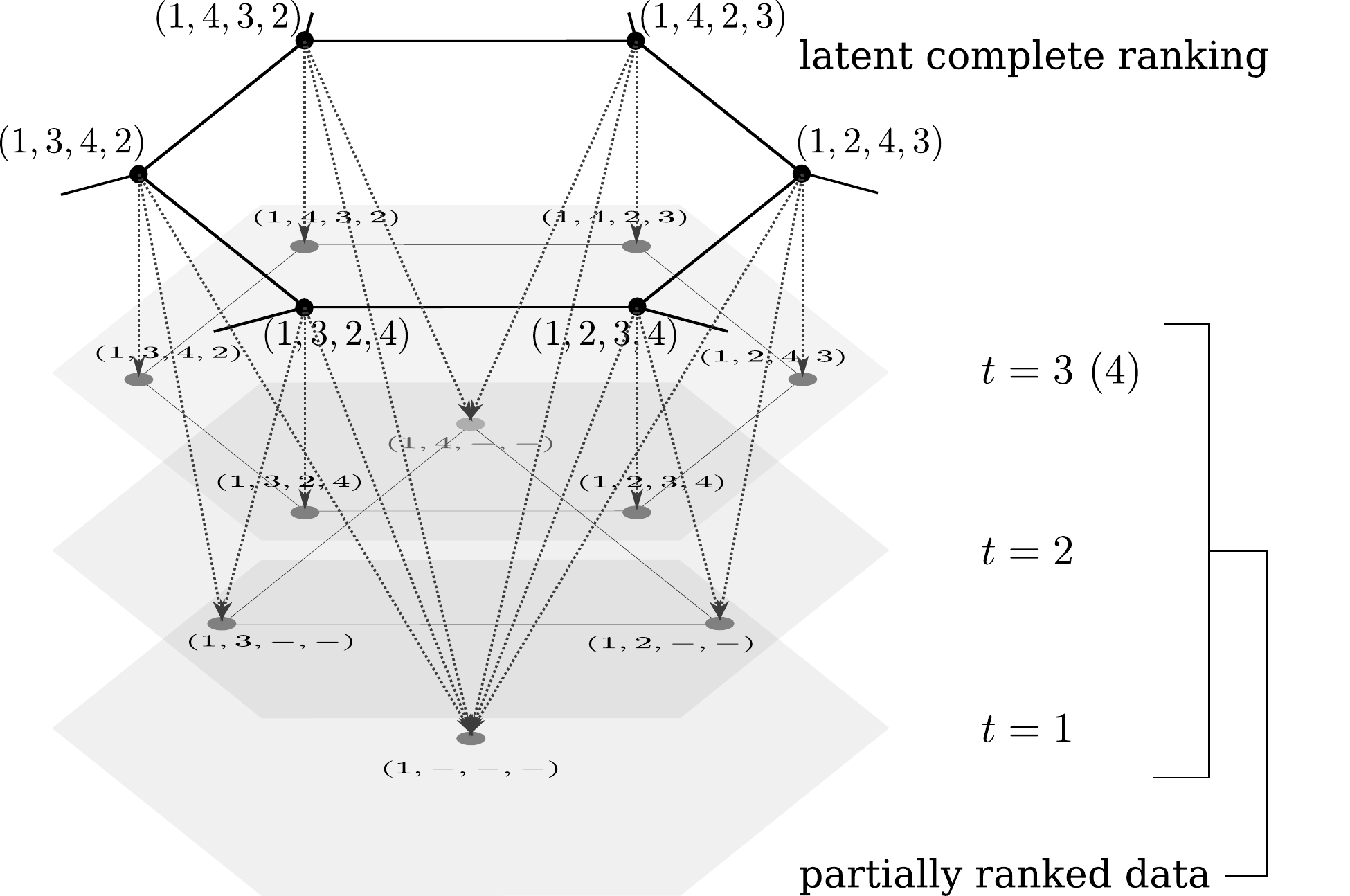}
    \caption{
    A latent structure behind partially ranked data when the number of items is four:
    A ranking is expressed as a list of ranked items.
    The number located at the $i$-th position of a list represents the label of the $i$-th preference.
    The top layer shows latent complete rankings with a graph structure.
    A vertex in the top layer corresponds to a latent complete ranking.
    A edge in the top layer is endowed by a distance between complete rankings.
    The bottom three layers show partial rankings generated according to missing mechanisms.
    An arrow from the top layer to the bottom three layers corresponds to a missing pattern.
    A probability on arrows from a complete ranking to the resulting partial rankings corresponds to a missing mechanism.
    }
    \label{fig:concept}
\end{figure}

Our ideas for the construction of the estimators are two-fold:
First, 
we work with a latent structure behind partially ranked data (see Figure \ref{fig:concept}).
This structure consists of the graph representing complete rankings (in the top layer)
and arrows representing missing patterns.
In this structure,
a vertex in the top layer represents a latent complete ranking;
An edge is endowed by a distance between complete rankings;
An arrow from the top layer to the bottom layers represents 
a missing pattern;
A multinomial distribution on arrows from a complete ranking
corresponds to a missing mechanism.
Second, 
we assume that two missing mechanisms become more similar as the associated complete rankings get closer to each other on the graph (in the top layer).
Together with both the restriction to the probability simplex and the EM algorithm, 
these ideas are implemented by the graph regularization method (\cite{hallac2015network}) under the probability restriction.
In addition, we discuss the convergence properties of the proposed method.

The simulation studies as well as applications to real data 
demonstrate that 
the proposed method improves on the existing methods under non-ignorable missing mechanisms,
and
the performance of the proposed method is comparable to those of the existing methods
under the MAR assumption.

\subsection{Literature review}
\label{section: literature}

Relatively scarce is the literature on the inference for ranking-related data with (non-ignorable) missing data.
\cite{Marlin:2007:CFM:3020488.3020521} points out that the MAR assumption does not hold in the context of the collaborative filtering.
\cite{Marlin:2007:CFM:3020488.3020521} and \cite{marlin2009collaborative} propose two estimators based on missing mechanisms. These estimators show the higher performance both in prediction of rating and in the suggestion of top-$t$ ranked items to users than estimators ignoring a missing mechanism.
Using the Plackett-Luce and the Mallows models,
\cite{fahandar2017statistical} introduces a rank-dependent coarsening model for 
pairwise ranking data.
This study is different from these studies in types of ranking-related data: 
\cite{marlin2009collaborative} and \cite{Marlin:2007:CFM:3020488.3020521} discuss rating data; \cite{fahandar2017statistical} discusses pairwise ranking data; this study discusses partially ranked data.

Several methods have been proposed
for estimating distributions of partially ranked data
(\cite{beckett1993maximum,busse2007cluster,meilua2010dirichlet,lebanon2008non,jacques2014model,caron2014bayesian}).
These methods regard partially ranked data as missing data.
 \cite{beckett1993maximum} discusses imputing items on missing positions of a partial ranking by employing the EM algorithm.
\cite{busse2007cluster} and \cite{meilua2010dirichlet} discuss 
the clustering of top-$t$ rankings
by the existing ranking distances for top-$t$ rankings.
\cite{lebanon2008non} proposes a non-parametric model together with a computationally efficient estimation method for partially ranked data. 
For the proposal, \cite{lebanon2008non} exploits the algebraic structure of partial rankings and utilizes the Mallows distribution as a smoothing kernel.
\cite{jacques2014model} proposes a clustering algorithm for multivariate partially ranked data. 
\cite{caron2014bayesian} discusses Bayesian non-parametric inferences of top-$t$ rankings on the basis of the Plackett-Luce model. 
\cite{caron2014bayesian} does not explicitly rely on the framework that regards partially ranked data as the result of missing data; However, the model discussed in \cite{caron2014bayesian} is equivalent to that under the MAR assumption.
Overall, all previous studies rely on the MAR assumption, whereas
our study is the first attempt to estimate the distribution of partially ranked data with a (possibly) non-ignorable missing mechanism.

We work with the graph regularization framework called Network Lasso (\cite{hallac2015network}).
Network Lasso employs the alternating direction method of multipliers (ADMM; see \cite{boyd2011distributed}) to solve a wide range of regularization-related optimization problems of a graph signal that cannot be solved efficiently by generic optimization solvers.
In addition, Network Lasso has a desirable convergence property,
cooperates with distributed processing systems,
and has been applied to various optimization problems on a graph.
We present an application of Network Lasso to missing data analysis.
In the application, we coordinate Network Lasso with the probability simplex constraint and the EM algorithm.

\subsection{Organization}
The rest of this paper is organized as follows.
Section 2 formulates a probabilistic model of partially ranked data based on a missing mechanism, and introduces a distance-based graph structure for a complete ranking.
Section 3 proposes the regularized estimator for both a latent complete ranking and a missing mechanism.
We also discuss the convergence properties of the proposed estimation procedure.
Section 4 demonstrates the result of simulation studies and real data analysis.
Section 5 concludes the paper.
The concrete algorithm of the proposed estimation procedure 
and
The proof of the convergence property are 
provided in Appendices A and B, respectively.
\section{Preliminaries}

\subsection{Notation}
\label{section: notation}

We begin with introducing notations for analyzing partially ranked data.
In this paper,
we identify a complete ranking of $r$ items $\{1,\ldots,r\}$
with a permutation that maps each item $i\in\{1,\ldots,r\}$ uniquely to a corresponding rank $\{1,\ldots,r\}$.
A top-$t$ ranking is a list of $t$ items out of $r$ items.
We identify a top-$t$ ranking with a permutation that maps an item in a subset of items uniquely to a corresponding rank in $\{1,\ldots,t\}$.

We denote by $S_{r}$ the collection of all complete rankings of $r$ items.
We denote by $\overline{S}_r$ the collection of all top-$t$ rankings with $t$ running through $t=1,\ldots,r-1$.
We denote by $t(\tau)$ the length of a partial ranking $\tau\in\overline{S}_{r}$.
A tuple of a complete ranking $\pi$ and length $t$ uniquely determines a top-$t$ ranking $\tau$:
$\pi^{-1}(i)=\tau^{-1}(i)$ for $i=1,\ldots,t$,
where $\pi^{-1}$ and $\tau^{-1}$ denote the inverses of $\pi$ and $\tau$, respectively.
We define the collection of complete rankings compatible with a given ranking $\tau\in \overline{S}_{r}$ as 
\[
[\tau] = \{\pi\in S_r \mid \pi^{-1}(i) = \tau^{-1}(i)\ (i=1,\ldots,t(\tau))\}.
\]

\subsection{Partial rankings and a missing mechanism}
\label{section: missing}
Next we introduce
notations and terminologies for a probabilistic model with a missing mechanism.

A probabilistic model for top-$t$ ranked data with a general missing mechanism consists of 
a probabilistic model of generating complete rankings
and that of a missing mechanism.
The joint probability of a complete ranking and a missing mechanism is decomposed as
\[
P(t,\pi)=P(t\mid \pi)P(\pi),
\]
where $P(\pi)$ determines how a complete ranking is filled,
and 
$P(t \mid \pi)$ specifies a missing pattern conditioned on the latent complete ranking $\pi$.
Then, the probability of a top-$t$ ranking $\tau\in\overline{S}_{r}$ is obtained by marginalizing a latent complete ranking out:
\[
P(\tau)= \sum_{\pi\in[\tau]}P(t(\tau)\mid \pi)P(\pi).
\]
Now distributions $P(t\mid\pi)$ and $P(\pi)$ are parameterized by $\phi$ and $\theta$,
respectively;
hence $P(\tau;\theta,\phi)=\sum_{\pi\in[\tau]}P(t\mid\pi;\phi)P(\pi;\theta)$.
We call $\{P(\pi;\theta):\theta\in\Theta\}$ with a parameter space $\Theta$ a complete ranking model
and $\{P(t\mid\pi;\phi):\phi\in\Phi\}$
with a parameter space $\Phi$ a missing model.
We call $\theta$ a complete ranking parameter, 
and call $\phi$ a missing parameter,
respectively.
Given the i.i.d.~observations $\tau_{(n)}=\{\tau_{1},\ldots,\tau_{n} \in \overline{S}_{r}\}$, we denote the negative log-likelihood function by
\begin{align}
L(\theta,\phi;\tau_{(n)}) := -\sum_{i=1}^n \log\left[\sum_{\pi\in[\tau_i]}P(t(\tau_i)\mid \pi;\phi)P(\pi;\theta)\right]. \label{likelihood}
\end{align}

\subsection{Distances, graph structures, and distributions on complete rankings}
\label{section: distance}

We end this section with introducing distances, graph structures, and distributions on complete rankings.

We endow the class $S_{r}$ of complete rankings with a distance structure as follows.
Since we identify the class $S_{r}$ as a the class of permutations,
we endow $S_{r}$ with the symmetric group structure and its group law $\circ$.
Using this identification, we leverage distances on symmetric groups for distances on $S_{r}$.
There exist a large number of distances on $S_{r}$ such as the Kendall distance,
the Spearman rank correlation metric, and the Hamming distance.
Among these, the Kendall distance (\cite{kendall1938new}) has been often used in statistics and machine learning.
The Kendall distance between two complete rankings $\pi_1$ and $\pi_2$, $d(\pi_{1},\pi_{2})$, is defined as
\[
d(\pi_{1},\pi_{2}):= \min_{\tilde{d}}\{\tilde{d}: \pi_{2}=a_{\tilde{d}}\circ\ldots\circ a_{1}\circ \pi_{1},\ a_{1},\ldots,a_{\tilde{d}}\in\mathcal{A}\},
\]
where $\mathcal{A}$ is the whole class of adjacent transpositions.
This distance is suitable for describing similarity between preferences because
the transform of a complete ranking by a single adjacent transposition is just the exchange of the $i$-th and $(i+1)$-th preferences.
In this paper, we focus on the Kendall distance $d$ as a distance on $S_{r}$.

There exists a one-to-one correspondence between the distance structure with the Kendall distance and a graph structure.
Set the vertex set $V=S_r$ and set the edge set $E=\{\{\pi,\pi'\}:\text{ there exists } b\in\mathcal{A} \text{ such that }\pi=b\circ \pi'\}$.
Then the distance structure $(S_{r},d)$ corresponds one-to-one to the graph $G=(V,E)$,
since the Kendall distance $d(\pi,\pi')$ is its minimum path length between the vertices corresponding to $\pi$,$\pi'$.
Remark that $E$ is rewritten as $E=\{\{\pi,\pi'\}: d(\pi,\pi')=1 \}$.

We introduce a well-known probabilistic model for complete rankings.
The Mallows model (\cite{mallows1957non}) is one of the most popular probabilistic models for complete rankings.
The Mallows model associated with the Kendall distance is defined as
\[
\left\{P(\pi;\sigma,c) = \frac{\exp\{-c d(\pi,\sigma)\}}{Z(c)}: c>0,\sigma\in S_{r}\right\},
\]
where $\sigma$ is a location parameter indicating a representative ranking, $c$ is a concentration parameter indicating a decay rate, and $Z(c)=\sum_{\pi\in S_r}\exp\{-c d(\pi,\sigma)\}$ is a normalizing constant that depends only on $c$. 
The mixture model of $K\in\mathbb{N}$ Mallows distributions is defined as
\begin{align}
\left\{P(\pi;\bm{c},\bm{\sigma},\bm{w}) = \sum_{k=1}^K w_{k} \frac{\exp\{-c_k d(\pi,\sigma_k)\}}{Z(c_{k})}: c_{k}>0,\ \sigma_{k}\in S_{r},\ w_{k}> 0,\ \sum_{k=1}^{K}w_{k}=1\right\} \label{mallows_mixture}
\end{align}
where $\bm{c} = \{c_k\}_k,\bm{\sigma} = \{\sigma_k\}_k,\bm{w} = \{w_k\}_k$ represent the parameters of each mixture component.
The Mallows mixture model has been used for estimation and clustering analysis of ranked data (\cite{murphy2003mixtures,busse2007cluster}).

\section{Proposed Method}
\label{section: method}

In this section, we propose estimators for both complete ranking and missing models together with a simple estimation procedure.
Here we assume that the parameterization of the complete ranking and missing models is separable, that is, $\theta$ and $\phi$ are distinct.
We use the following missing model $\{P(t\mid \pi; \phi):\phi\in\Phi\}$ to allow a non-ignorable missing mechanism:
\begin{align*}
    P(t\mid \pi; \phi) &= \phi_{\pi, t} \text{ for } t\in\{1,\ldots,r-1\} \text{ and } \pi\in S_{r},\\
    \Phi&=\left\{\phi \in \mathbb{R}^{r!(r-1)}: \sum_{t=1}^{r-1}\phi_{\pi, t} = 1, \phi_{\pi,t}\geq 0, \pi \in S_{r}\right\}.
\end{align*}
We make no assumptions on a complete ranking model $\{P(\pi; \theta):\theta\in\Theta\}$.

\subsection{Estimators and estimation procedure}
\label{section: procedure}
We propose the following estimators for $\theta$ and $\phi$:
On the basis of i.i.d.~observations $\tau_{(n)}=\{\tau_{1},\ldots,\tau_{n} \in \overline{S}_{r}\}$,
\[
(\hat{\theta}(\tau_{(n)}),\hat{\phi}(\tau_{(n)}))=\argmin_{\theta\in\Theta, \phi\in\Phi} L_{\lambda}(\theta,\phi;\tau_{(n)}).
\]
Here $L_{\lambda}$ with a regularization parameter $\lambda>0$ is defined as
\[
L_{\lambda}(\theta,\phi;\tau_{(n)})= L(\theta,\phi;\tau_{(n)}) +\lambda\sum_{\{\pi,\pi'\}\in E}\|\phi_{\pi}-\phi_{\pi'}\|_2^2,
\]
where 
$\phi_{\pi}$ with $\pi\in S_{r}$ denotes the vector $(\phi_{\pi, 0},\ldots,\phi_{\pi, (r-1)})$,
and recall that
$L(\theta,\phi;\tau_{(n)})$ is the negative log-likelihood function (\ref{likelihood})
and
$E$ is the edge set of the graph induced by the Kendall distance;
see Subsections \ref{section: missing}-\ref{section: distance}.

We conduct minimization in the definition of $\hat{\theta},\hat{\phi}$ using the following EM algorithm: 
At the $(m+1)$-th step, set 
\begin{align}
    \theta^{m+1}&=\argmin_{\theta'\in\Theta} L(\theta';\tau_{(n)},q^{m+1}_{(n)}), \label{opttheta}\\
    \phi^{m+1}&=\argmin_{\phi' \in\Phi} L_{\lambda}(\phi';\tau_{(n)},q_{(n)}^{m+1}), \label{optphi}
\end{align}
where for $i\in\{1,\ldots,n\}$,
\begin{align}
q_{i,\pi}^{m+1}&=
\begin{cases}
\phi^{m}_{\pi,t(\tau_{i})}P(\pi;\theta^{m})
\bigg{/}\sum_{\pi'\in[\tau_i]}\phi^{m}_{\pi',t(\tau_{i})}P(\pi';\theta^{m}) & (\pi\in[\tau_i]), \\
0 & (\pi\not\in[\tau_i]),
\end{cases}\label{optq}\\
q_{(n)}^{m+1}&:=\{q_{i,\pi}^{m+1}: i=1,\ldots,n , \pi\in S_{r} \},\\
L(\theta;\tau_{(n)},q^{m+1}_{(n)})&:=-\sum_{i=1}^n\sum_{\pi\in[\tau_i]}q^{m+1}_{i,\pi}\log P(\pi;\theta),\\
L_{\lambda}(\phi;\tau_{(n)},q^{m+1}_{(n)})&:=-\sum_{i=1}^n\sum_{\pi\in[\tau_i]}q^{m+1}_{i,\pi}\log\phi_{\pi', t(\tau_i)} +\lambda \sum_{\{\pi,\pi'\}\in E}\|\phi_{\pi}-\phi_{\pi'}\|_2^2. \label{likelihood_phi}
\end{align}

Consider minimizations (\ref{opttheta}) and (\ref{optphi}).
Minimization (\ref{opttheta}) depends on the form of a complete ranking model $P(\pi;\theta)$;
For example,
consider the Mallows model with $\theta=(\sigma,c)$ (see Section \ref{section: distance}).
In this case, we write down the minimization of $\theta$ at the $(m+1)$-th step as follows:
\begin{align*}
\sigma^{m+1}&=\mathop{\argmin}_{\tilde\sigma\in S_r}\sum_{i=1}^n\sum_{\pi\in S_r}q_{i,\pi}^{m+1} d(\pi,\tilde\sigma),\\
c^{m+1}&=\argmin_{\tilde{c}>0}\sum_{i=1}^n\sum_{\pi\in S_r}q_{i,\pi}^{m+1}\{\tilde{c}d(\pi,\sigma^{m+1}) + \log(Z(\tilde{c}))\}.
\end{align*}
See \cite{busse2007cluster} for more details.
Minimization (\ref{optphi}) in the $(m+1)$-th step is conducted using the following iteration:
At the $(l+1)$-th step, set
\begin{align}
\phi^{l+1}&=\argmin_{\tilde{\phi}\in\Phi }L_{\rho}(\tilde{\phi},\varphi^{l},u^{l}; q^{m+1}_{(n)}),\label{optvertex}\\
\varphi^{l+1}&=\argmin_{\tilde{\varphi}\in\mathbb{R}^{r!(r-1)^2}}L_{\rho}(\phi^{l+1},\tilde{\varphi},u^{l}; q^{m+1}_{(n)}),\label{optvarphi}\\
u^{l+1}&=u^{l}+(\phi^{l+1}-\varphi^{l+1}). \label{optu}
\end{align}
Here, $\varphi\in \mathbb{R}^{r!(r-1)^2}$ is the copy variable of $\phi$, $u\in \mathbb{R}^{r!(r-1)^2}$ is the dual variable, and $L_\rho$ is an augmented Lagrangian function with a penalty constant $\rho$ defined as
\begin{align}
L_{\rho}(\phi,\varphi,u;q^{m+1}_{(n)})&=-\sum_{\pi\in V}\sum_{t}[q^{m+1}_{\pi, t}\log \phi_{\pi, t}] \nonumber\\
&\quad+ \sum_{\{\pi,\pi'\}\in E}\left\{\lambda\|\varphi_{\pi, \pi'}-\varphi_{\pi',\pi}\|^2_2-\frac{\rho}{2}(\|u_{\pi, \pi'}\|_2^2+\|u_{\pi',\pi}\|_2^2)\right.\nonumber\\
&\quad\quad\quad\quad\quad\left.+\frac{\rho}{2}(\|\phi_{\pi}-\varphi_{\pi, \pi'}+u_{\pi, \pi'}\|_2^2+\|\phi_{\pi'}-\varphi_{\pi',\pi}+u_{\pi',\pi}\|_2^2)\right\},\label{lagrangian}
\end{align}
where 
\begin{align}
q^{m+1}_{\pi, t}:=\sum_{i:t(\tau_i)=t}q_{i,\pi}^{m+1},  \label{qpit}
\end{align}
for all $\pi\in S_{r}$ and $t=1,\ldots,r-1$.
Note that $q^{m+1}_{\pi,t}=0$ when $\{i:t(\tau_i)=t\}$ is an empty set.
The detailed algorithm is provided in Appendix \ref{section: algorithm}.

\begin{remark}[Meaning of the penalty term]
We make the assumption that two complete rankings close to each other in the Kendall distance have smoothly related missing probabilities.
This assumption leads to adding a ridge penalty
\[
\sum_{\substack{\pi,\pi'\in S_r\\d(\pi,\pi')=1}}\|\phi_{\pi}-\phi_{\pi'}\|_2^2
=\sum_{\{\pi,\pi'\}\in E}\|\phi_{\pi}-\phi_{\pi'}\|_{2}^{2}
\]
to the negative log-likelihood function.
This assumption is reasonable because the Kendall distance measures the similarity of preferences expressed by two rankings.
\end{remark}

\begin{remark}[The proposed method under the MAR assumption]
For top-$t$ ranked data, the MAR assumption is expressed as
\[
P(t(\tau)\mid\pi;\phi) = P(t(\tau)\mid\pi';\phi),\ (\pi,\pi' \in [\tau]).
\]
Then under the MAR assumption, the negative log-likelihood function $L(\theta,\phi;\tau_{(n)})$ is decomposed as
\begin{align*}
L(\theta,\phi;\tau_{(n)})&=-\sum_{i=1}^n\log P(\tau_i\mid\pi\in[\tau_i];\phi)-\sum_{i=1}^n\log \left[\sum_{\pi\in[\tau_i]}P(\pi;\theta)\right]\nonumber\\
&=L(\phi;\tau_{(n)})+L(\theta;\tau_{(n)}),
\end{align*}
which indicates that the parameter estimation for $\phi$ is unnecessary for estimating $\theta$.
\end{remark}

\subsection{Convergence}
In this subsection,
we discuss theoretical guarantees for two procedures (\ref{opttheta})-(\ref{likelihood_phi}) and (\ref{optvertex})-(\ref{optu}).

It is guaranteed that 
the sequence $\{L_{\lambda}(\theta^m,\phi^m;\tau_{(n)}):m=1,2,\ldots \}$ obtained using the procedure (\ref{opttheta})-(\ref{likelihood_phi}) monotonically decreases,
\[L_{\lambda}(\theta^{m},\phi^{m};\tau^{(n)}) \geq L_{\lambda}(\theta^{m+1},\phi^{m+1};\tau^{(n)}), \ m=1,2,\ldots,\]
because the procedure is just the EM algorithm.
Introduce a latent assignment variable $z_{(n)}=\{z_i\}_i$ ($z_i\in \mathbb{R}^{|S_r|}$ for every $i = 1,\ldots,n$. $z_{i\pi}=1$ if $\tau_i$ is the missing from $\pi$ and $z_{i\pi}=0$ otherwise).
Using $z_{(n)}$, we decompose the 
likelihood function as follows:
\begin{align*}
    &L_\lambda(\theta,\phi;\tau_{(n)},z_{(n)}) \\
    &= -\sum_{i=1}^n\log\left[\prod_{\pi\in[\tau_i]}\{\phi_{\pi,  t(\tau_i)}P(\pi;\theta)\}^{z_{i,\pi}}\right] +\lambda\sum_{\{\pi,\pi'\}\in E}\|\phi_{\pi}-\phi_{\pi'}\|_2^2.\\
    &=\left\{-\sum_{i=1}^n\sum_{\pi\in[\tau_i]}z_{i,\pi}\log P(\pi;\theta)\right\}+\left\{-\sum_{i=1}^n\sum_{\pi\in[\tau_i]}z_{i,\pi}\log\phi_{\pi,  t(\tau_i)} +\lambda\sum_{\{\pi,\pi'\}\in E}\|\phi_{\pi}-\phi_{\pi'}\|_2^2\right\}\\
    &=L(\theta;\tau_{(n)},z_{(n)})+L_{\lambda}(\phi;\tau_{(n)},z_{(n)}).
\end{align*}
On the basis of the decomposition,
the standard procedure of the EM algorithm yields the iterative algorithm shown in (\ref{opttheta})-(\ref{likelihood_phi}).
Note that it depends on a complete ranking model $P(\pi;\theta)$ 
whether the convergent point of the sequence is a local maximum of $L(\theta,\phi;\tau_{(n)})$;
see Section 3 of \cite{mclachlan2007algorithm}.

Next, it is guaranteed that the sequence $\phi^{l+1}$ obtained using the procedure (\ref{optvertex})-(\ref{optu}) converges to the global minimum of $L_{\lambda} (\phi ; \tau_{(n)} , q_{(n)}^{m} )$ in the sense of $L_{\lambda}(\phi;\tau_{(n)},q^{m}_{(n)})$.
\begin{proposition}
\label{prop: convergence_admm}
 The sequence $\{L_{\lambda}(\phi^l,\tau_{(n)},q^{m+1}_{(n)})\}_{l=1}^{\infty}$ converges to $\min_{\phi}L_{\lambda}(\phi,\tau_{(n)},q^{m+1}_{(n)})$.
\end{proposition}

The proof is provided in Appendix \ref{section: proof}.
The basis of the proof is reformulating the optimization problem (\ref{optphi}) as an instance of the alternating direction method of multipliers (ADMM; \cite{boyd2011distributed}:
We rewrite the problem (\ref{optphi}) as follows:
\begin{eqnarray}
\phi=&\argmin_{\phi'}&-\sum_{\pi\in V}\sum_{t}[q_{\pi, t}\log \phi'_{\pi, t}] +\lambda \sum_{\{\pi,\pi'\}\in E}\|\phi_{\pi}-\phi_{\pi'}\|_2^2\nonumber\\
&&\mathrm{s.t.}\ \sum_t \phi_{\pi, t} = 1 \ (\forall \pi\in V) \label{RFopt}
\end{eqnarray}
where $V$ is the vertex set of the graph defined in Section \ref{section: distance} and $q_{\pi, t}=\sum_{i:t(\tau_i)=t}\sum_{k=1}^Kq_{i,k,\pi}$.
Introducing a copy variable $\varphi$ on the edge set,
we recast the optimization problem (\ref{RFopt}) into an equivalent form:
\begin{eqnarray}
\phi,\varphi=&\argmin_{\phi',\varphi'}&-\sum_{\pi\in V}\sum_{t}[q_{\pi, t}\log \phi'_{\pi, t}] + \lambda\sum_{\{\pi,\pi'\}\in E}\|\varphi'_{\pi, \pi'}-\varphi'_{\pi',\pi}\|^2_2\label{optphidecomp}\\
&\mathrm{s.t.}&\phi'_{\pi}=\varphi'_{\pi, \pi'} \ (\forall \{\pi,\pi'\}\in E),\nonumber\\
&& \sum_t \phi'_{\pi, t} = 1 \ (\forall \pi\in V).\nonumber
\end{eqnarray}
Note that this reformulation follows the idea of \cite{hallac2015network}.
We employ ADMM to solve
the optimization of the sum of objective functions of splitted variables under linear constraints.

\section{Numerical experiments}

In this section,
we apply our methods to both simulation studies and real data analysis. 
In simulation studies, we use the Mallows mixture models (\ref{mallows_mixture}) with two types of missing models.
In the real data analysis, we use the election records of the American Psychological Association collected in 1980.

\subsection{Performance measures}
 
 We evaluate the performance of several estimators for $\theta$ and $\phi$ in estimating distributions of a latent complete ranking and of a partial ranking.
 We measure the performance using the following total variation losses:
 When the true values of a complete ranking and missing parameters are
 $\theta$ and $\phi$, respectively,
 the losses of estimators $\hat{\theta}$ and $\hat{\phi}$ are given as
\begin{align}
    L_{\mathrm{par}}
    &=L_{\mathrm{par}}(\theta,\phi;\hat{\theta},\hat{\phi})
    =\sum_{\tau\in \overline{S}_{r}}
    |P(\tau;\theta,\phi)-P(\tau;\hat{\theta},\hat{\phi})| \label{Lpar}\\
    L_{\mathrm{comp}}
    &=L_{\mathrm{comp}}(\theta,\hat{\theta})
    =\sum_{\pi\in S_{r}}
    |P(\pi;\theta)-P(\pi;\hat{\theta})|.\label{Lcomp}
\end{align}
Losses $L_{\mathrm{par}}$ and $L_{\mathrm{comp}}$ measure the estimation losses for partial and complete ranking distributions, respectively.

\subsection{Method comparison}
\label{section: comparison methods}

We compare our estimators with the estimator based on the maximum entropy approach proposed by \cite{busse2007cluster} and a non-regularized estimator, abbreviated by ME and NR, respectively.
In addition,
we use the proposed estimator with the regularization parameter selected using two-fold cross-validation based on $L_{\mathrm{par}}$.

We denote the proposed method introduced in section \ref{section: method} with the value of regularization parameter $\lambda$ as R$\lambda$ and that with the regularization parameter selected using cross-validation as RCV.

The maximum entropy approach (ME) uses an extended distance between top-$t$ rankings to introduce an exponential family distribution of a top-$t$ ranking.
From the viewpoint of missing data analysis, ME implicitly assumes the MAR assumption.
For this reason,
in the maximum entropy approach,
we estimate the missing model parameter $\phi$ by assuming homogeneous missing probabilities $P(t\mid \pi) = \phi_t\ (\forall \pi\in S_r)$ and using the maximum likelihood in the evaluation of the loss $L_{\mathrm{par}}(\theta,\phi)$.

The non-regularized estimator (NR) is the minimizer of the non-regularized likelihood function $L(\theta,\phi; \tau_{(n)})$.
The estimation based on the non-regularized likelihood function can be implemented straightforwardly.

\subsection{Stopping criteria}
In simulation studies,
we use the following stopping criteria and hyperparameters.
We terminate the iteration of the EM algorithm 
when 
the change of the likelihood of the observable distribution gets lower than $\epsilon = 1$.
We terminate the iteration of ADMM when both the primal and dual residuals got less than $\epsilon_p=\epsilon_d=1$
or when the number of iterations exceeded 100.
In addition,
to prevent being trapped in local minima due to the EM algorithm,
we use the following devices.
First, we use 10 different initial location parameters in the EM algorithm.
Second,
we make the value of the location parameter transit from the current to a different one in the first five iterations of the EM algorithm.

\subsection{Simulation studies}
\label{section: synthetic}

We conducted two simulation studies.
The data-generating models are as follows:
For complete ranking models,
we use the Mallows and Mallows mixture models.
For missing models,
we use a binary missing mechanism in which the possible missing patterns are only that no items are missing or that all but the first items are missing.
We parameterize missing models in such a way that there is a discrepancy between the distribution of a complete ranking generated by the latent Mallows model and the marginal distribution of a partial ranking restricted to $S^{(r-1)}_{r}:=\{\tau:t(\tau)=r-1,\tau\in \overline{S}_{r}\}$.
Note that $S^{(r-1)}_{r}$ is identical to $S_{r}$ as a set.

In each simulation,
we generate 100 datasets with sample size of $n=1000$.
We set the number of items to $r=5$.

\subsubsection{Tilting the concentration parameter}
\label{section: tilt concentration}
\begin{figure}[htbp]

 \begin{minipage}{0.32\hsize}
  \begin{center}
   \includegraphics[width=45mm]{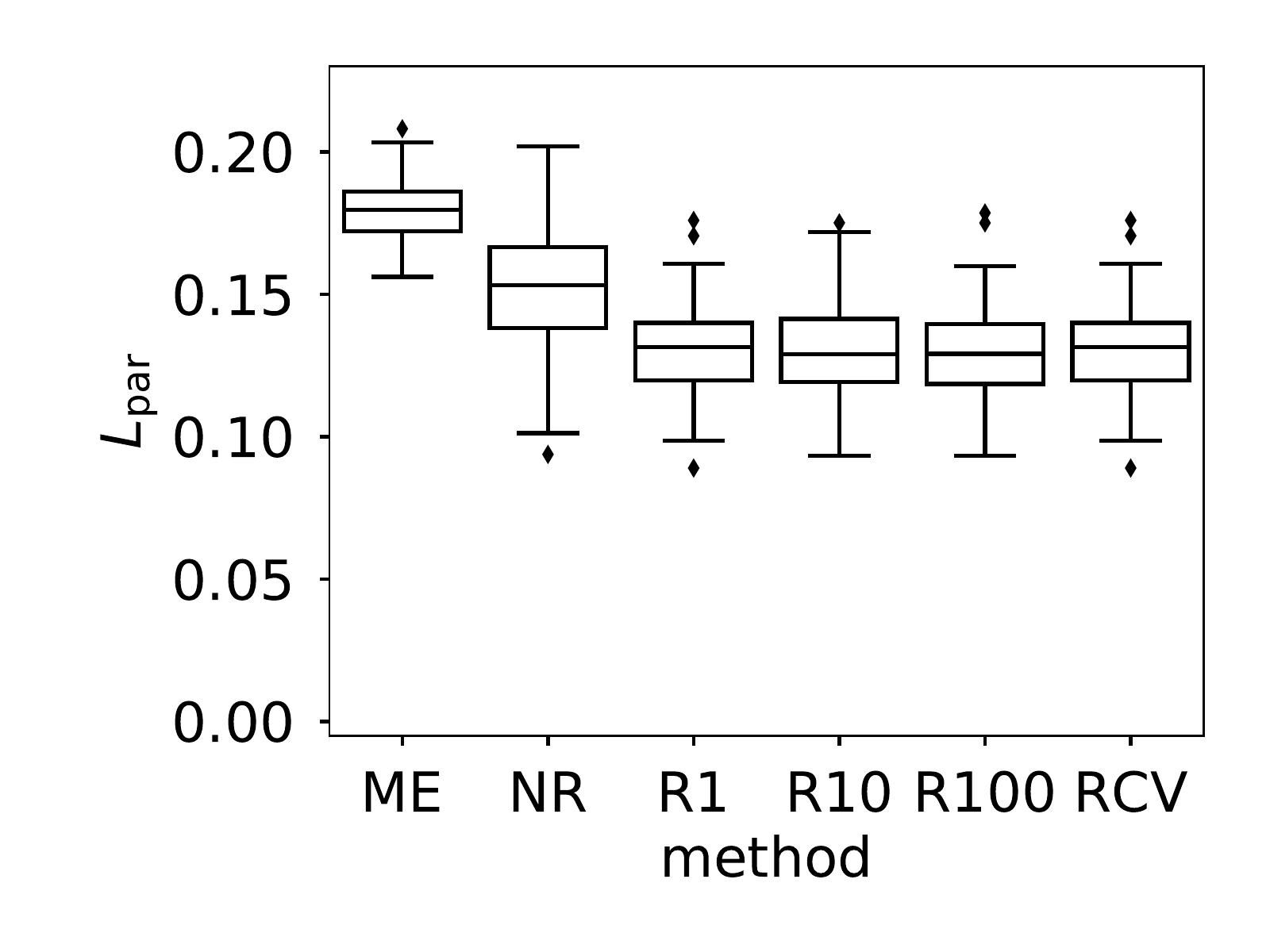}
  \end{center}
  \subcaption{$c^{\ast}=0.8$}
  \label{p8}
 \end{minipage}
  \begin{minipage}{0.32\hsize}
  \begin{center}
   \includegraphics[width=45mm]{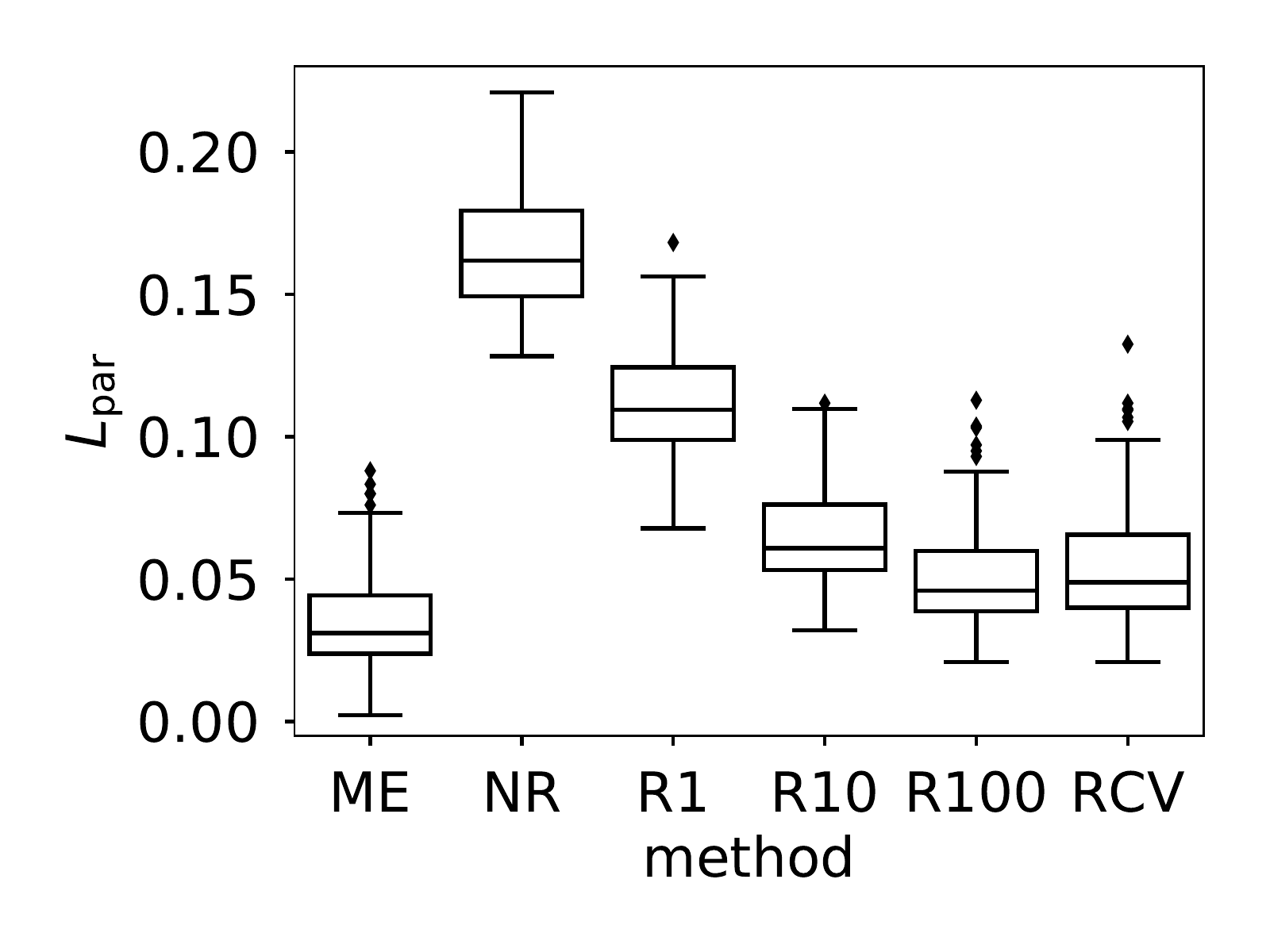}
  \end{center}
  \subcaption{$c^{\ast}=1.0$}
  \label{p10}
 \end{minipage}
  \begin{minipage}{0.32\hsize}
  \begin{center}
   \includegraphics[width=45mm]{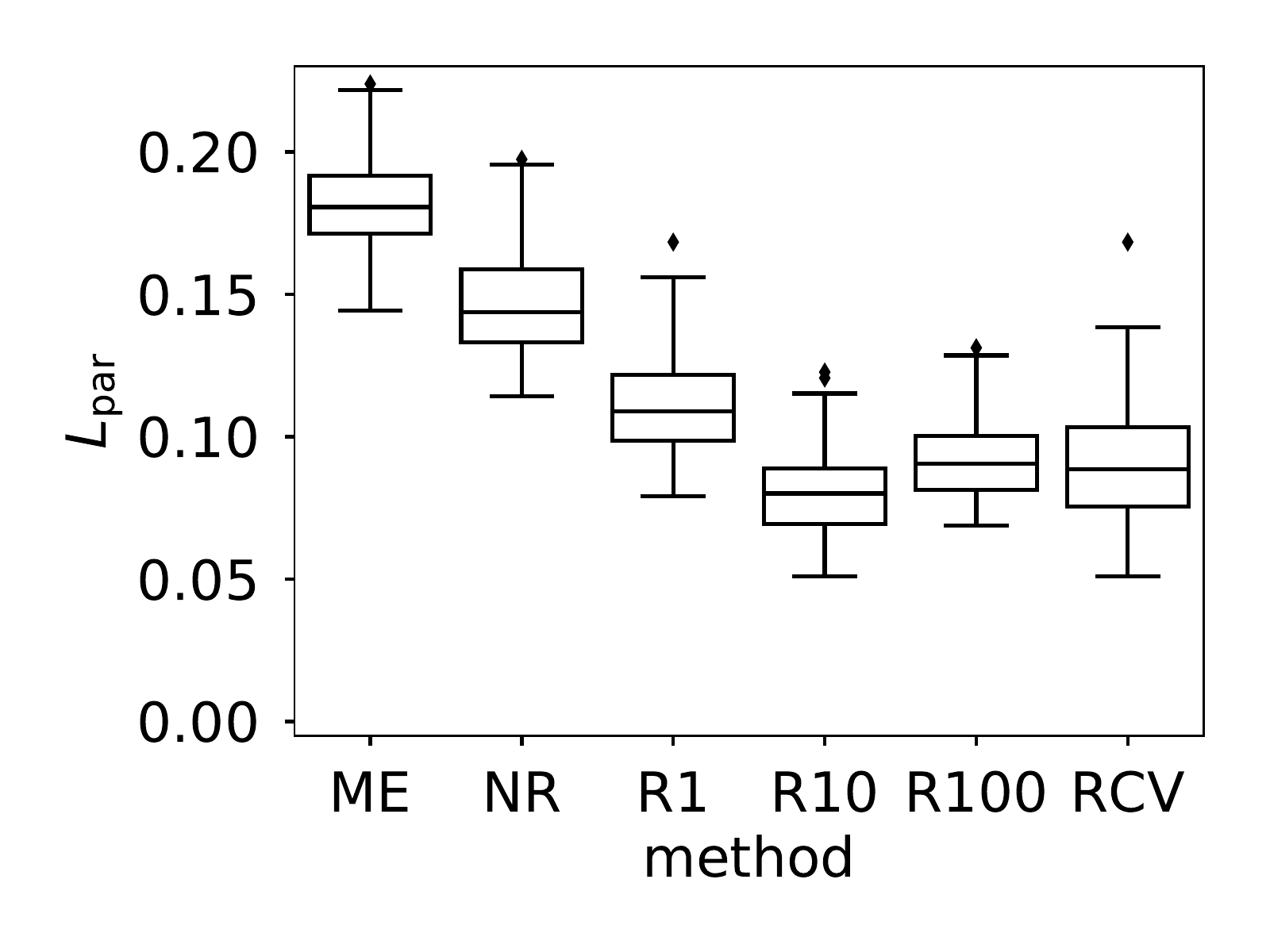}
  \end{center}
  \subcaption{$c^{\ast}=1.2$}
  \label{p12}
 \end{minipage}
 \caption{Boxplots of $L_{\mathrm{par}}$ in (\ref{Lpar}) for the dataset tilting the concentration parameter: The compared methods are the maximum entropy approach (ME), the non-regularized estimator (NR), the proposed methods (R1; R10; R100), and its version with cross-validation (RCV).
 The results with different values of the concentration parameter $c^{\ast}$ are shown in (A)-(C).}
 \label{synGP}
\end{figure}

\begin{figure}[htbp]
 \begin{minipage}{0.32\hsize}
  \begin{center}
   \includegraphics[width=50mm]{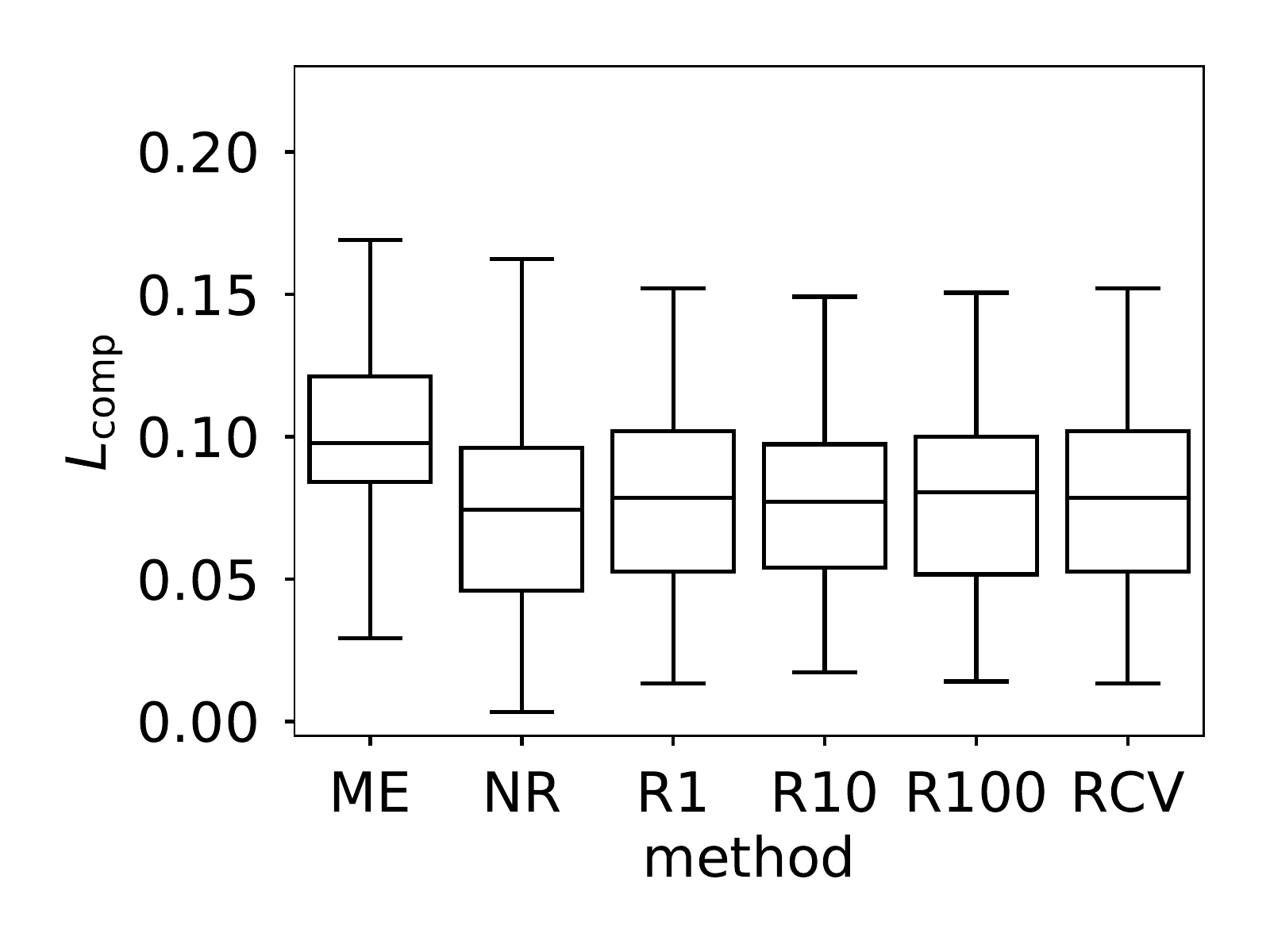}
  \end{center}
  \subcaption{$c^{\ast}=0.8$}
  \label{fig:one}
 \end{minipage}
  \begin{minipage}{0.32\hsize}
  \begin{center}
   \includegraphics[width=50mm]{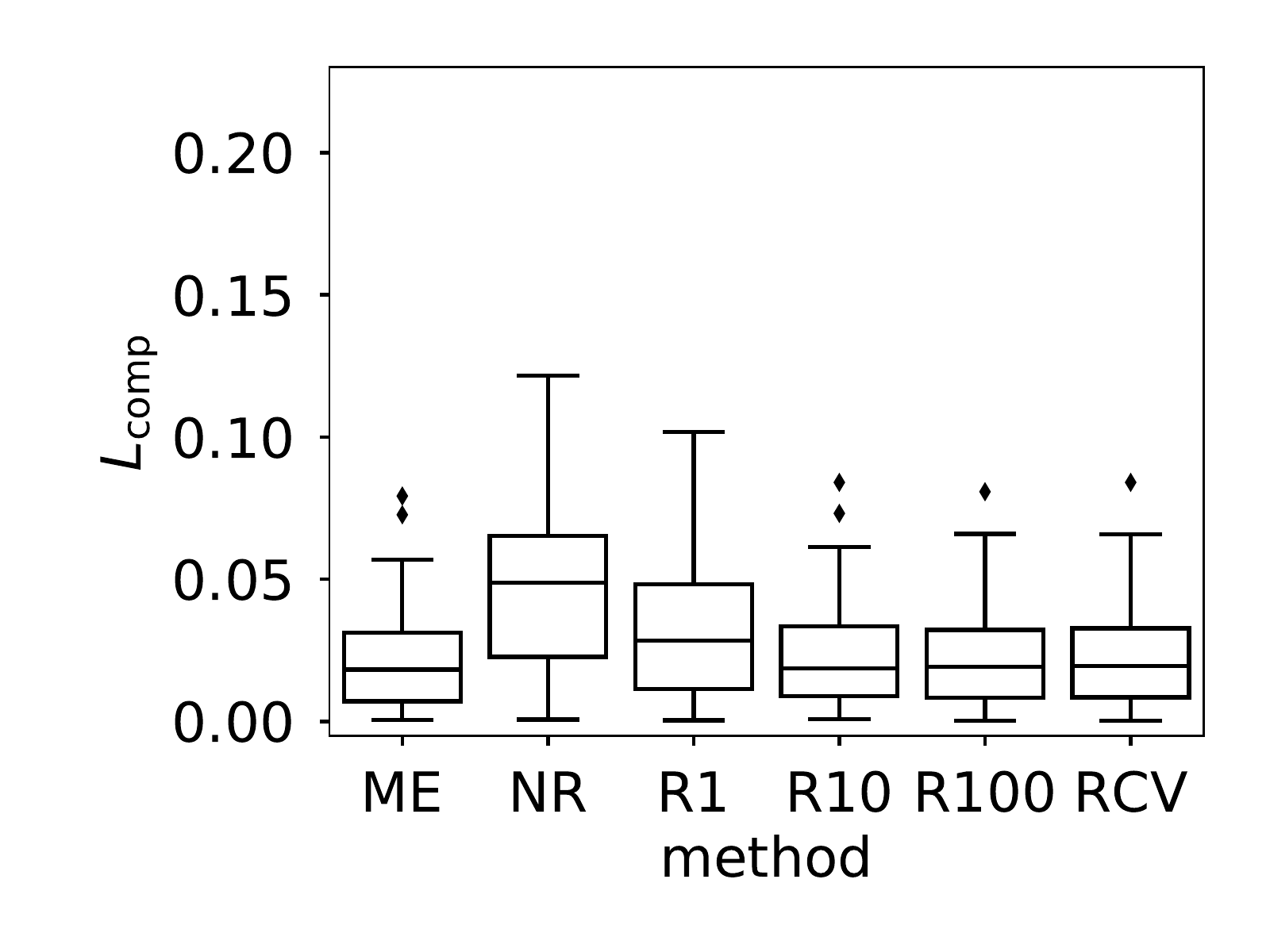}
  \end{center}
  \subcaption{$c^{\ast}=1.0$}
  \label{fig:one}
 \end{minipage}
  \begin{minipage}{0.32\hsize}
  \begin{center}
   \includegraphics[width=50mm]{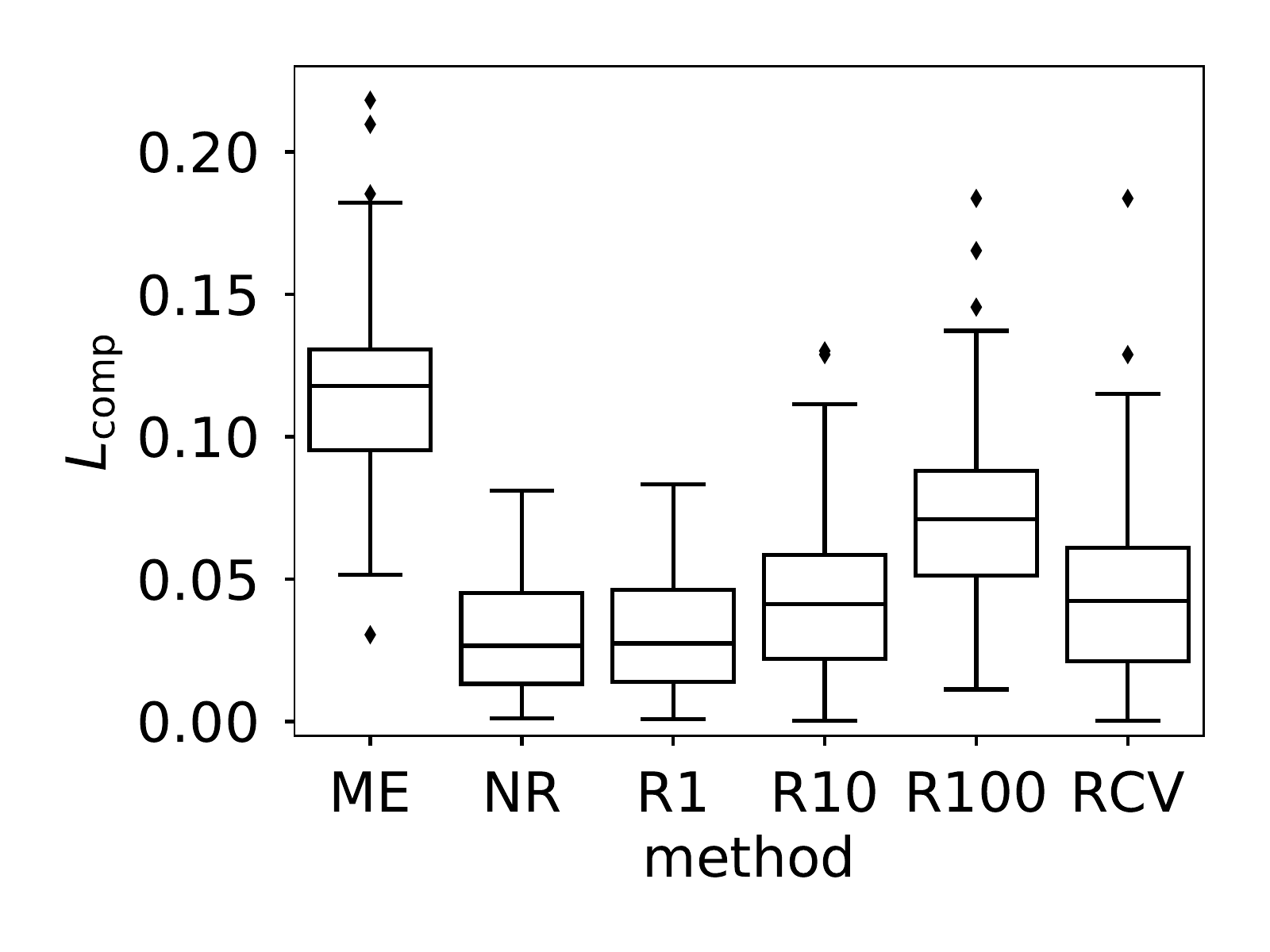}
  \end{center}
  \subcaption{$c^{\ast}=1.2$}
  \label{fig:one}
 \end{minipage}
 \caption{Boxplots of $L_{\mathrm{comp}}$ in (\ref{Lcomp}) for the dataset tilting the concentration parameter: The compared methods are the maximum entropy approach (ME), the non-regularized estimator (NR), the proposed methods (R1; R10; R100), and its version with cross-validation (RCV).  
 The results with different values of the concentration parameter $c^{\ast}$ are shown in (A)-(C).}
 \label{synGC}
\end{figure}
In the first simulation study,
we use the Mallows model and the missing model that tilts the concentration parameter $c$: 
The missing model is parameterized by $c^{\ast}>0$ and $R\in [0,1]$ as
\[
\phi_{\pi} = ( \phi_{\pi,0},\ldots,\phi_{\pi,(r-1)})
= (1-C_{\pi}(c^{\ast},R),0,\ldots,0,C_{\pi}(c^{\ast},R)), \ \pi\in S_{r}
\]
where
\[
C_{\pi}(c^{\ast},R) =\min\left\{1,\frac{Z(c)}{Z(c^{\ast})}R\exp \{-(c^{\ast}-c) d(\pi,\sigma_0)\}\right\}.
\]
In this parameterization,
the parameter
$c^{\ast}$ specifies the degree of concentration of
the marginal distribution $P(\tau;\theta,\phi)$ 
of a partial ranking restricted to $S^{(r-1)}_{r}$:
If 
$\{Z(c)/Z(c^{\ast})\}R\exp \{-(c^{\ast}-c) d(\pi,\sigma_0)\}\leq 1$,
$P(\tau;\theta,\phi)$ has the form of the Mallows distribution with the concentration parameter $c^{\ast}$:
\begin{align*}
P(\tau;\theta,\phi) &= \sum_{\pi'\in[\tau]}P(t=r\mid \pi';\phi)P(\pi';\theta)\\
&=C_{\pi}(c^{\ast},R)\frac{1}{Z(c)}\exp\{-c d(\pi,\sigma)\}\\
&= \frac{R}{Z(c^{\ast})}\exp\{-c^{\ast}d(\pi,\sigma_0)\},
\end{align*}
where $\pi(i)=\tau(i), i=1,\ldots,r-1$.
The parameter $0\leq R\leq 1$ specifies the proportion of partial rankings in $S^{(r-1)}_r$.
We set $c = 1$, $R=0.7$, and $c^{\ast} \in \{0.8,1,1.2\}$.

 Figures \ref{synGP} and \ref{synGC} show the results.
 When $c^{\ast}\neq 1$, the proposed methods outperform ME both in $L_{\mathrm{par}}$ and $L_{\mathrm{comp}}$.
 When $c^{\ast}=1$, the proposed methods underperform compared to ME.
 These results reflect that the setting with $c^{\ast}=1$ satisfies the MAR assumption, whereas the settings with $c^{\ast}\neq 1$ do not satisfy the MAR assumption.
For $L_{\mathrm{par}}$, the proposed methods outperform NR regardless of the values of $c^{\ast}$.
However, there are subtle distinctions in the values of $L_{\mathrm{comp}}$ of these methods.
The performance of the proposed method with the cross-validated regularization parameter (RCV) is comparable with that of the proposed method with the optimal regularization parameter both for $L_{\mathrm{par}}$ and $L_{\mathrm{comp}}$, indicating the utility of cross-validation.

\subsubsection{Tilting the mixture coefficient}
\label{section: tilt mixture}
In the second simulation study, we use the Mallows mixture model with two clusters and the missing model that tilts the mixture coefficient $w$.
We instantiate a missing model, in which missing probabilities depend on the cluster assignment $k\in\{1,\ldots,K\}$, such that $P(t\mid \pi,z_k=1)=P(t\mid z_k=1)=\phi_{k,t}$, where $z_k=1$ if and only if the assigned cluster is $k$ and $z_k=0$ otherwise.
Then,
the missing model is parameterized by $w^{\ast}\in [0,1]$ and $R\in [0,1]$ as
\[
\phi = \left(
	\begin{array}{ccccc}
	\phi_{1,1} & \ldots \phi_{1,(r-1)}\\
	\phi_{2,1} & \ldots \phi_{2,(r-1)}
	\end{array}
\right)=\left(
	\begin{array}{ccccc}
	1-C_{1}(w^{\ast},R) & 0 &\ldots & 0 & C_{1}(w^{\ast},R)\\
	1-C_{2}(w^{\ast},R) & 0 &\ldots& 0 & C_{2}(w^{\ast},R)
	\end{array}
\right),
\]
where
$C_{k}(w^{\ast},R) = (w^{\ast}_k / w_k)R.$
In this parameterization, the parameter $w^{\ast}$ determines the mixture coefficient of the marginal distribution $P(\tau;\theta,\phi)$ of a partial ranking restricted to $S^{(r-1)}_r$.
We set the parameter values as follows:
\begin{itemize}
\item $\bm\sigma=((1,2,3,4,5), (3,2,5,4,1))$, $\bm{c} = (1,1)$, and $w = (0.5,0.5)$;
\item $R=0.7$ and $w^{\ast}=\{(0.5,0.5),(0.6,0.4),(0.7,0.3)\}$.
\end{itemize}

In this simulation study,
we additionally use the classification error as the performance measure.

\begin{figure}[htbp]
 \begin{minipage}{0.32\hsize}
  \begin{center}
   \includegraphics[width=50mm]{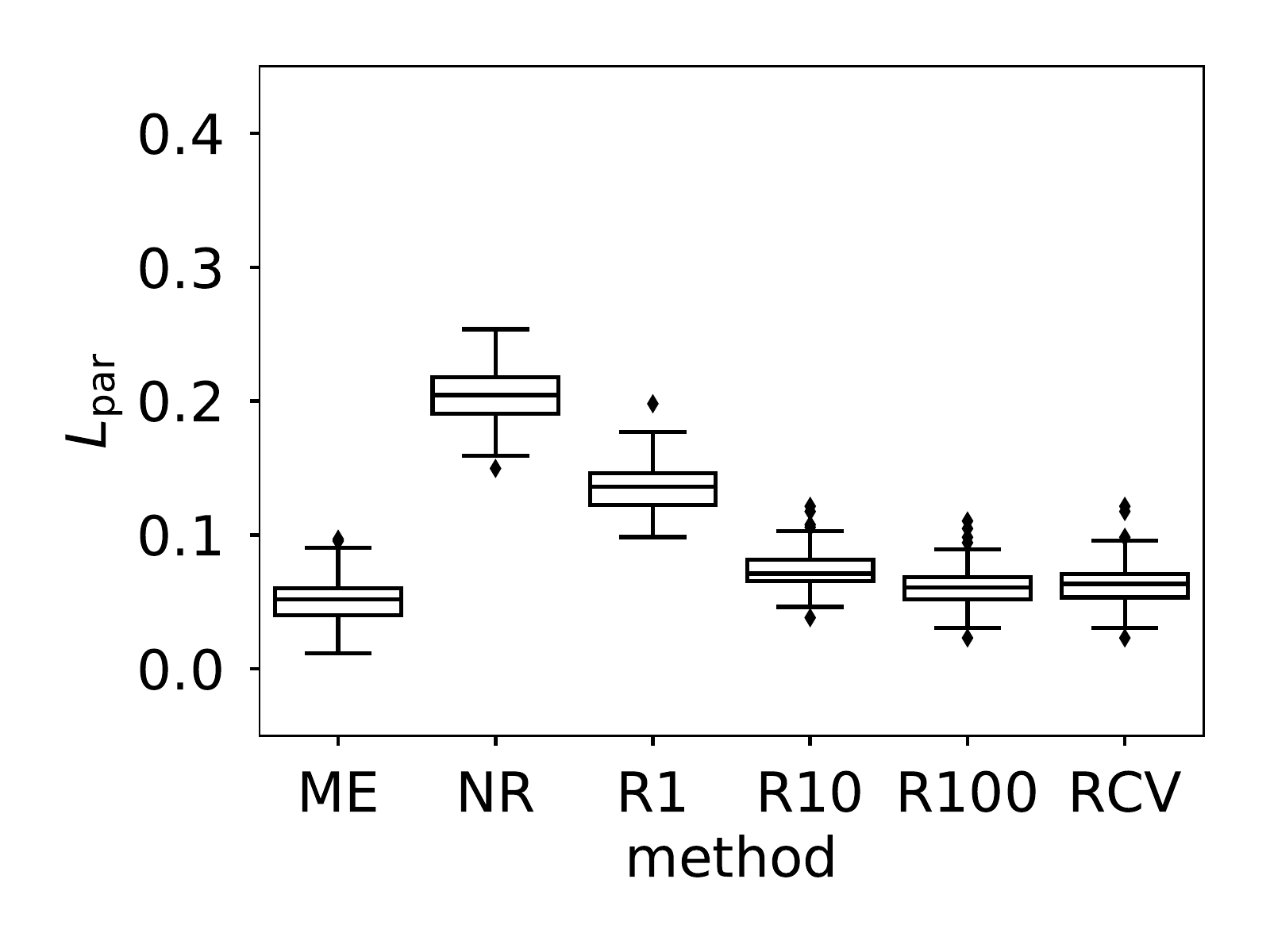}
  \end{center}
  \subcaption{$w^{\ast}_1=0.5$}
  \label{fig:one}
 \end{minipage}
  \begin{minipage}{0.32\hsize}
  \begin{center}
   \includegraphics[width=50mm]{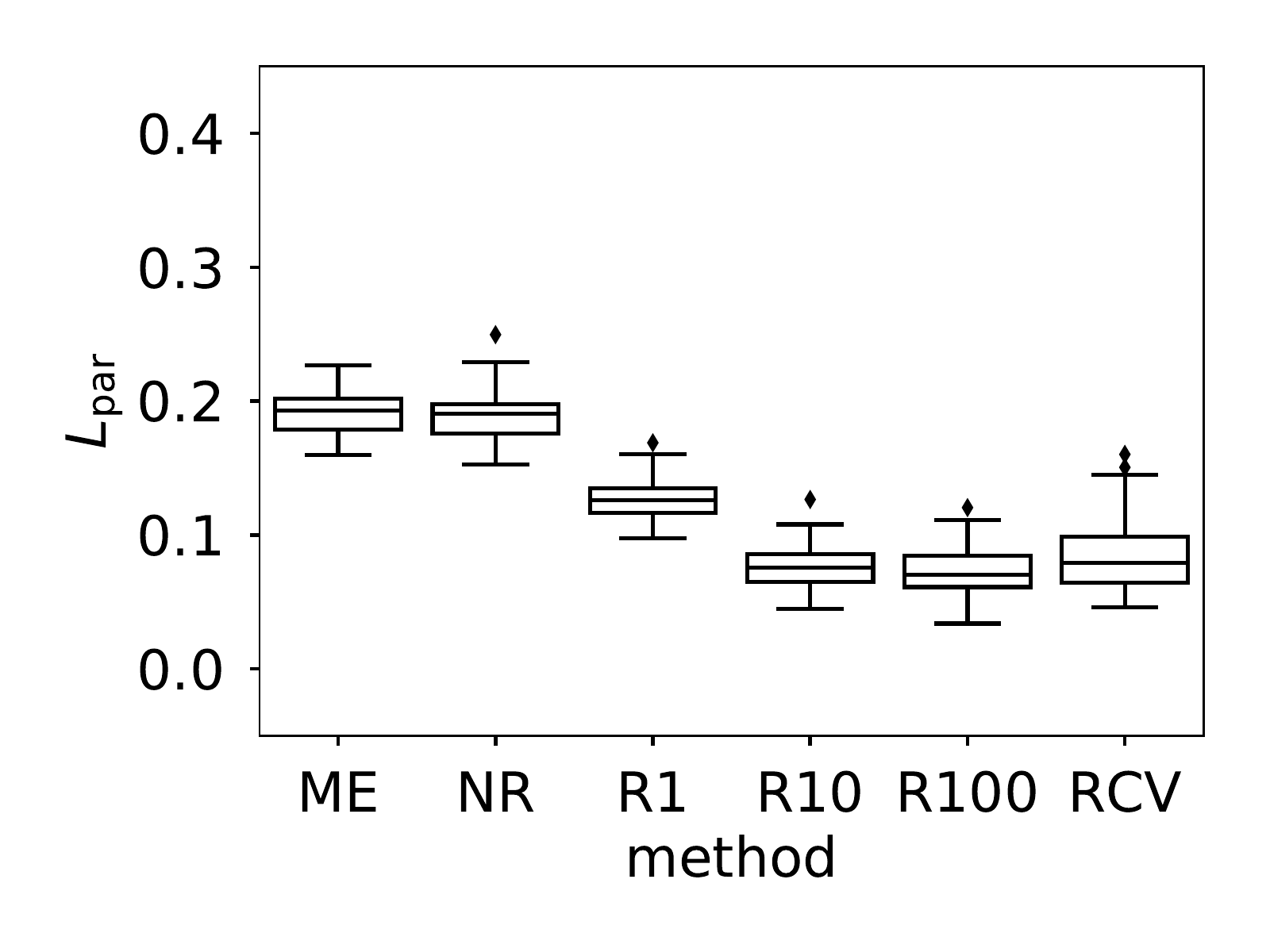}
  \end{center}
  \subcaption{$w^{\ast}_1=0.6$}
  \label{fig:one}
 \end{minipage}
  \begin{minipage}{0.32\hsize}
  \begin{center}
   \includegraphics[width=50mm]{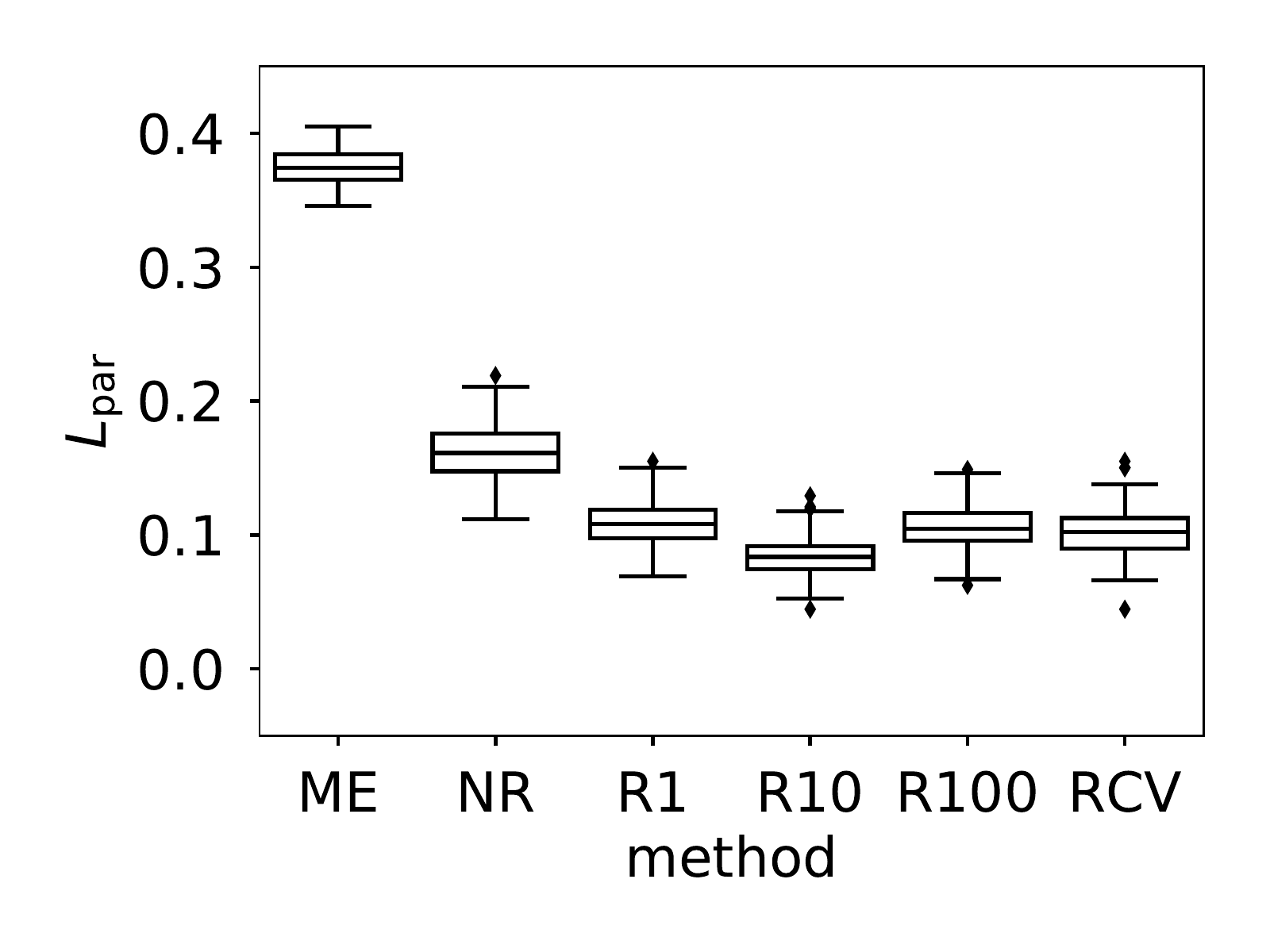}
  \end{center}
  \subcaption{$w^{\ast}_1=0.7$}
  \label{fig:one}
 \end{minipage}
 \caption{Boxplots of $L_{\mathrm{par}}$ in (\ref{Lpar}) for the dataset tilting the mixture coefficient: The compared methods are the maximum entropy approach (ME), the non-regularized estimator (NR), the proposed methods (R1; R10; R100) and its version with cross-validation (RCV).  The results with different values of the concentration parameter $w^{\ast}$ are shown in (A)-(C).}
 \label{synCP}
\end{figure}

\begin{figure}[htbp]
 \begin{minipage}{0.32\hsize}
  \begin{center}
   \includegraphics[width=50mm]{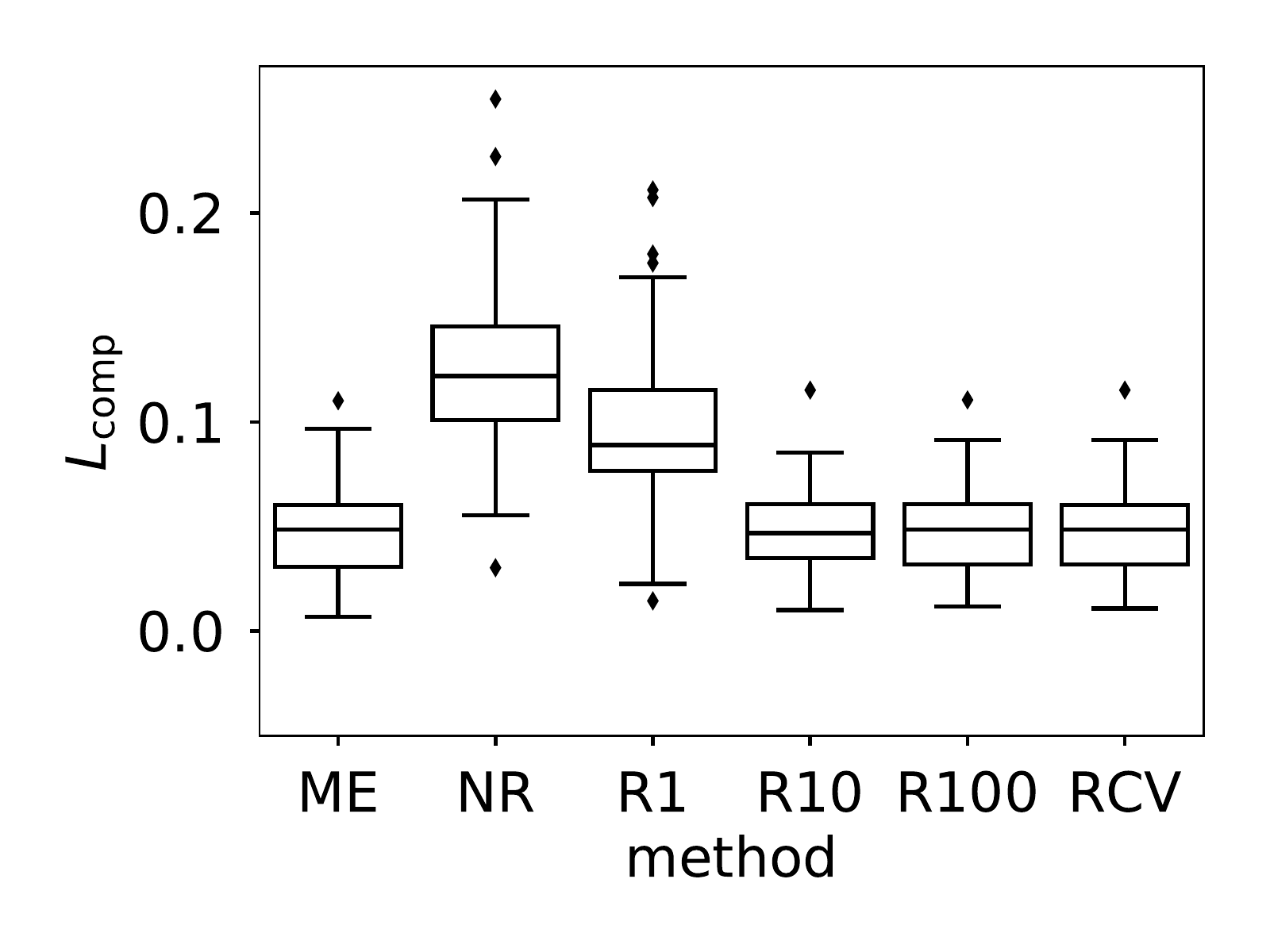}
  \end{center}
  \subcaption{$w^{\ast}_1=0.5$}
  \label{fig:one}
 \end{minipage}
  \begin{minipage}{0.32\hsize}
  \begin{center}
   \includegraphics[width=50mm]{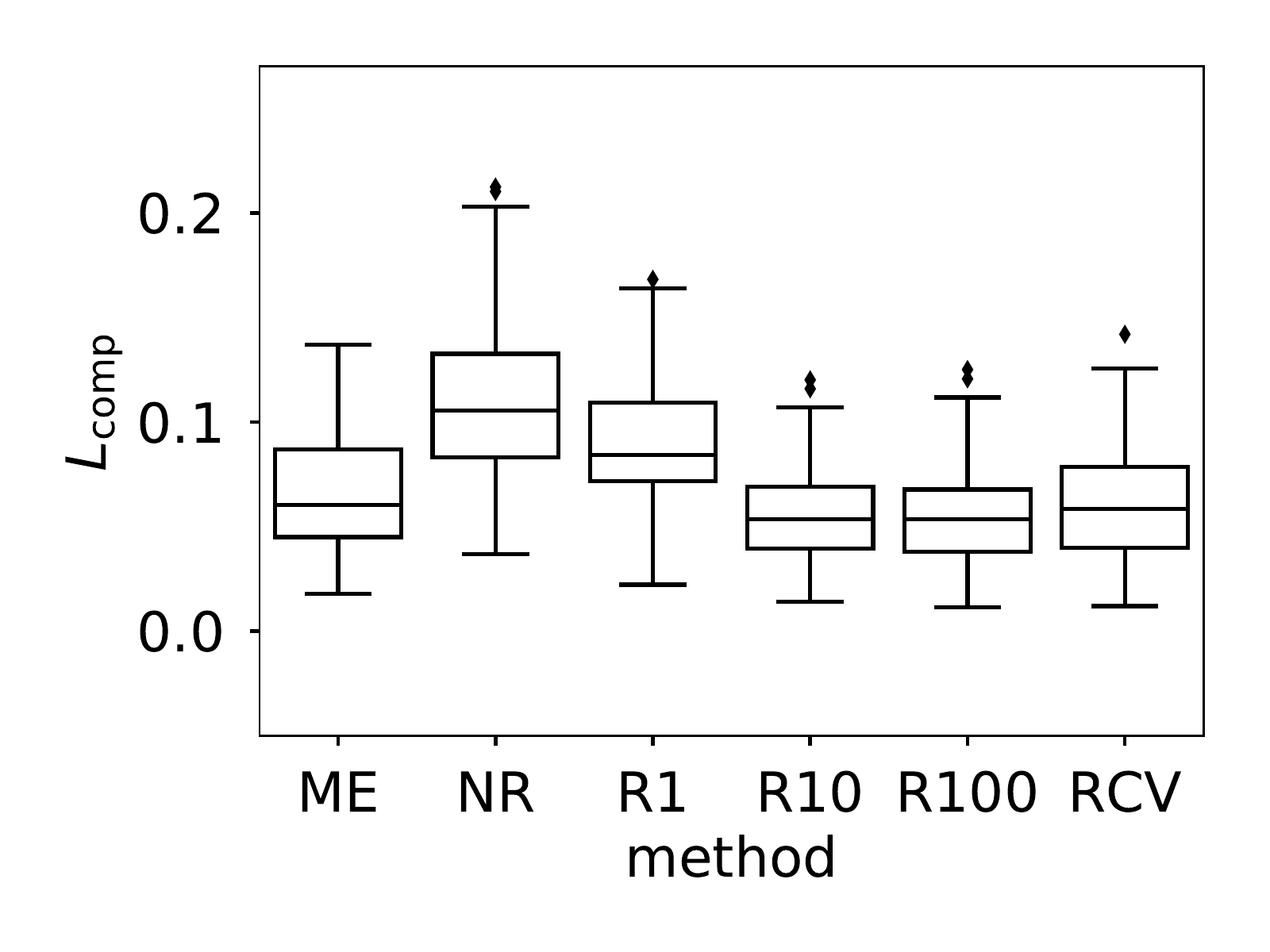}
  \end{center}
  \subcaption{$w^{\ast}_1=0.6$}
  \label{fig:one}
 \end{minipage}
  \begin{minipage}{0.32\hsize}
  \begin{center}
   \includegraphics[width=50mm]{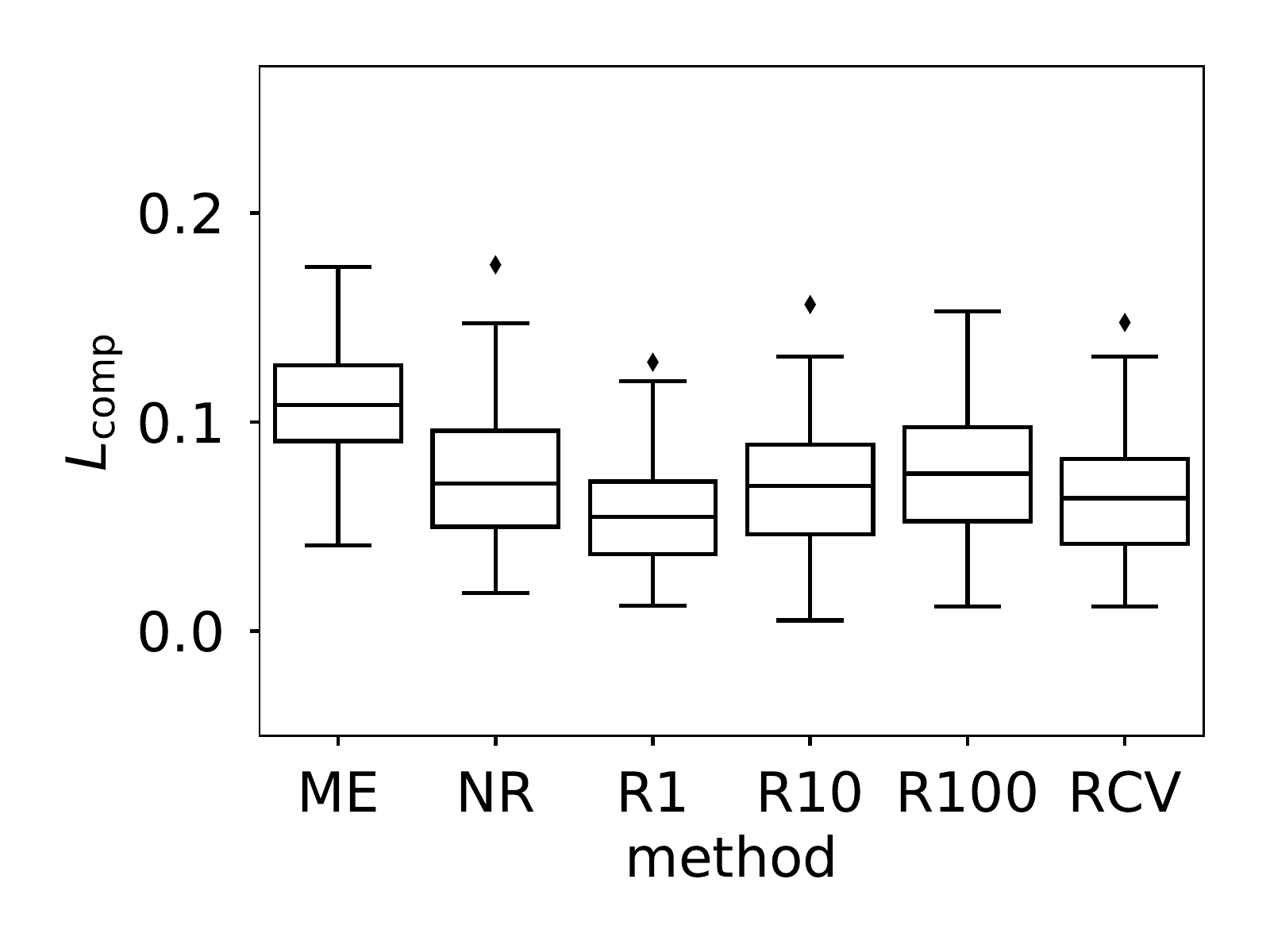}
  \end{center}
  \subcaption{$w^{\ast}_1=0.7$}
  \label{fig:one}
 \end{minipage}
 \caption{Boxplots of $L_{\mathrm{comp}}$ in (\ref{Lcomp}) for the dataset tilting the mixture coefficient: The compared methods are the maximum entropy approach (ME), the non-regularized estimator (NR), the proposed methods (R1; R10; R100) and its version with the cross-validation (RCV). The results with different values of the concentration parameter $w^{\ast}$ are shown in (A)-(C).}
 \label{synCC}
\end{figure}
\begin{figure}[H]
 \begin{minipage}{0.32\hsize}
  \begin{center}
   \includegraphics[width=50mm]{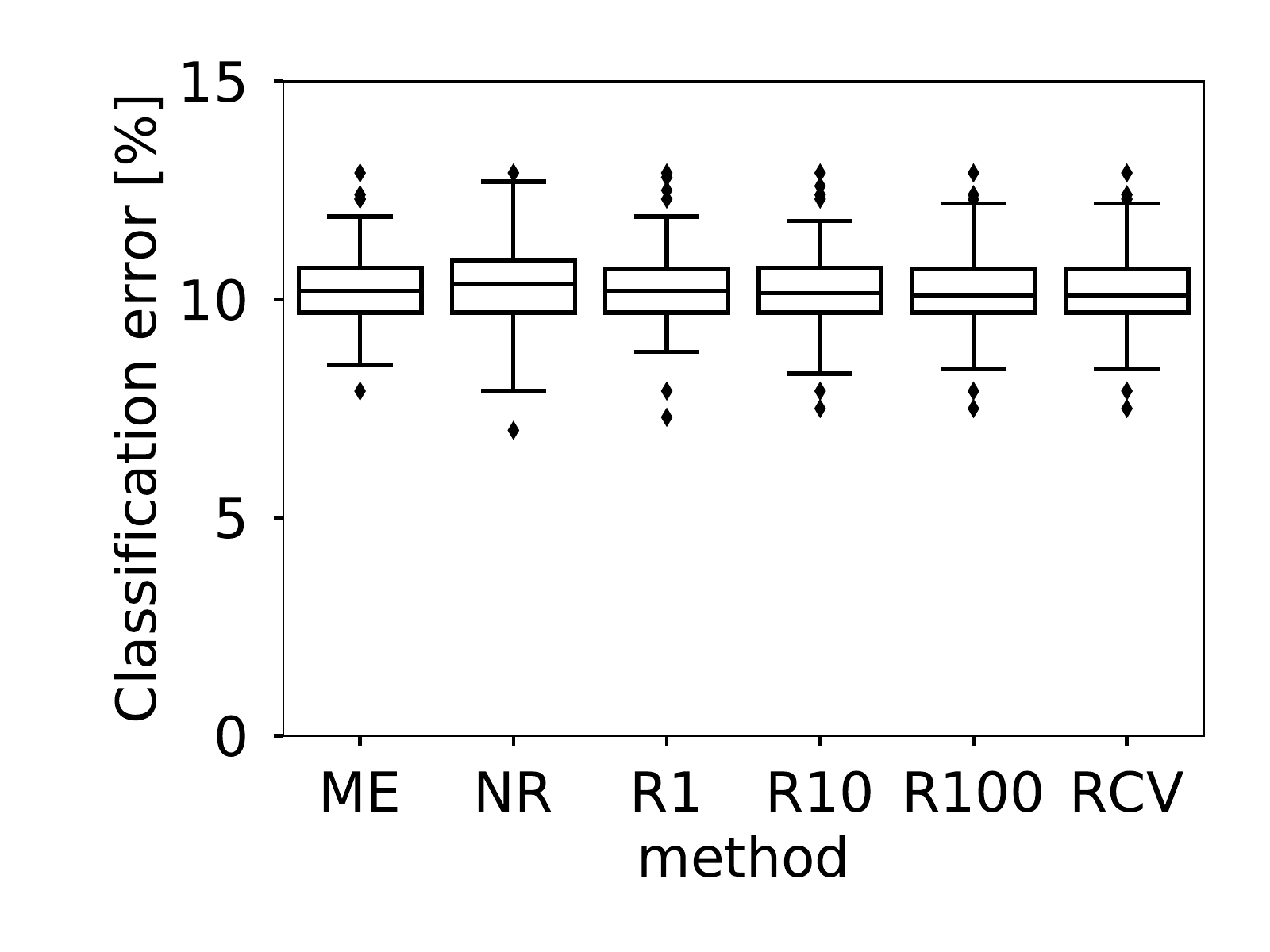}
  \end{center}
  \subcaption{$w^{\ast}_1=0.5$}
  \label{fig:one}
 \end{minipage}
  \begin{minipage}{0.32\hsize}
  \begin{center}
   \includegraphics[width=50mm]{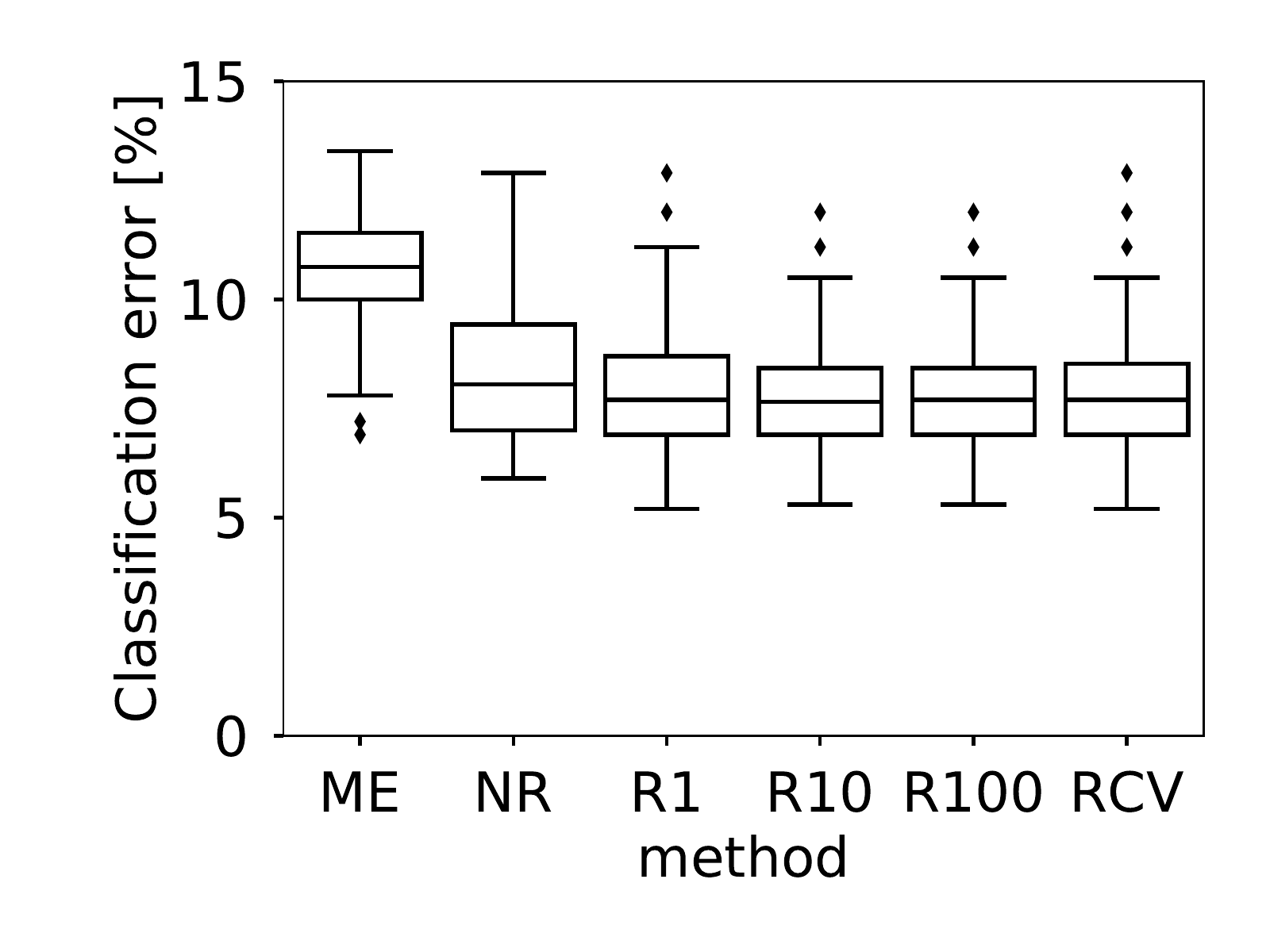}
  \end{center}
  \subcaption{$w^{\ast}_1=0.6$}
  \label{fig:one}
 \end{minipage}
  \begin{minipage}{0.32\hsize}
  \begin{center}
   \includegraphics[width=50mm]{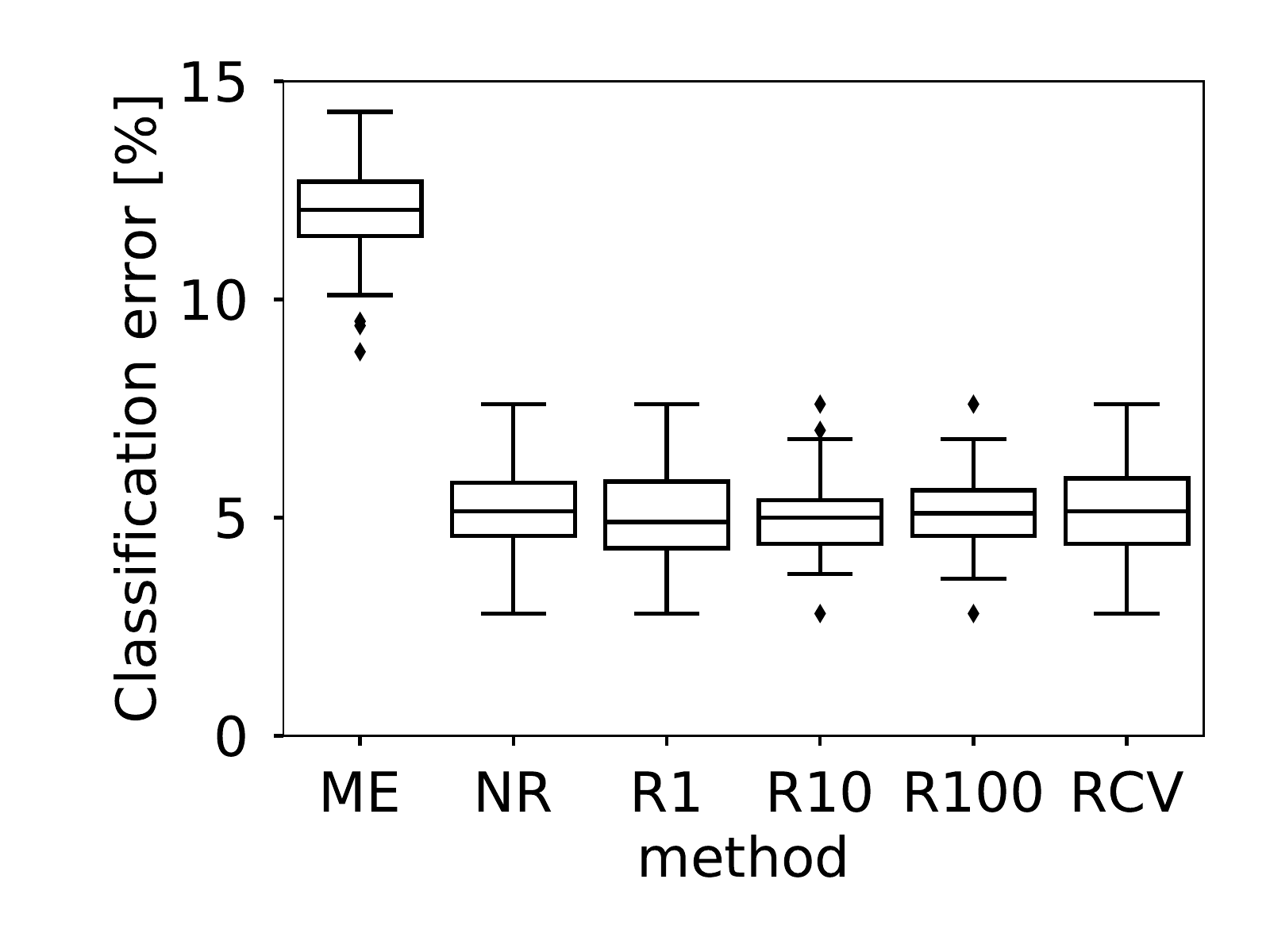}
  \end{center}
  \subcaption{$w^{\ast}_1=0.7$}
  \label{fig:one}
 \end{minipage}
 \caption{Boxplots of the classification errors for the dataset tilting the mixture coefficient: The compared methods are the maximum entropy approach (ME), the non-regularized estimator (NR), the proposed methods (R1; R10; R100) and its version with the cross-validation (RCV). The results with different values of the concentration parameter $w^{\ast}$ are shown in (A)-(C).}
 \label{synCD}
\end{figure}

Figures \ref{synCP}--\ref{synCD} show the results. 
The proposed methods outperform ME when $w_1^{\ast}\neq 0.5$ in comparing $L_{\mathrm{par}}$;
when $w_1^{\ast}$ is $0.7$ in comparing $L_{\mathrm{comp}}$;
when $w_1^{\ast}\neq 0.5$ in comparing the classification error.
The proposed methods outperform NR both in comparing $L_{\mathrm{par}}$ and $L_{\mathrm{comp}}$ except when $w_1^{\ast}$ is $0.7$ for $L_{\mathrm{comp}}$.
As $w^{\ast}$ deviates from $0.5$, the classification error of ME increases.
On the other hand, the classification errors of the other methods decrease.

\subsection{Application to real data}
We apply the proposed method to real data.
We use the election records for five candidates collected by the American Psychological Association. Among the 15549 vote casts, only 5141 filled all candidates ($t=5,4$); 2108 filled $t=3$; 2462 filled $t=2$; and the rest filled only $t=1$. 

\begin{figure}[H]
    \centering
    \includegraphics[width=50mm]{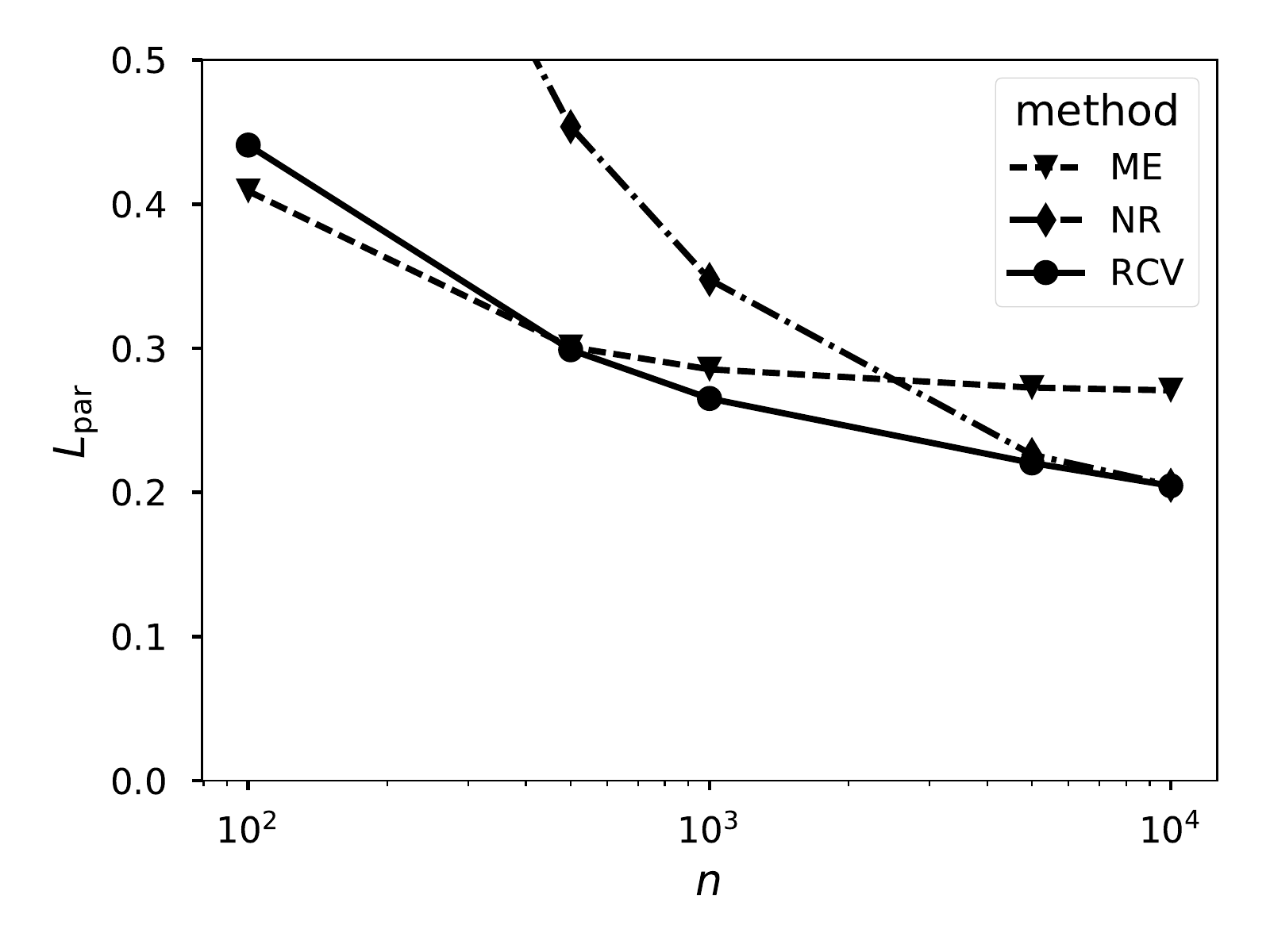}
    \caption{Sample size dependency of the mean of $L_{\mathrm{par}}$ in (\ref{Lpar}) for the American Psychological Association dataset: The vertical axis represents the mean value of $L_{\mathrm{par}}$; and the horizontal axis represents the sample size $n$ on the log-scale. The results with $n=100, 500, 1000, 5000, 10000$ are shown. The compared methods are the maximum entropy approach (ME), the non-regularized estimator (NR), the proposed method with cross-validation (RCV).}
    \label{apa2}
\end{figure}
\begin{figure}[H]
 \begin{minipage}{0.32\hsize}
  \begin{center}
   \includegraphics[width=50mm]{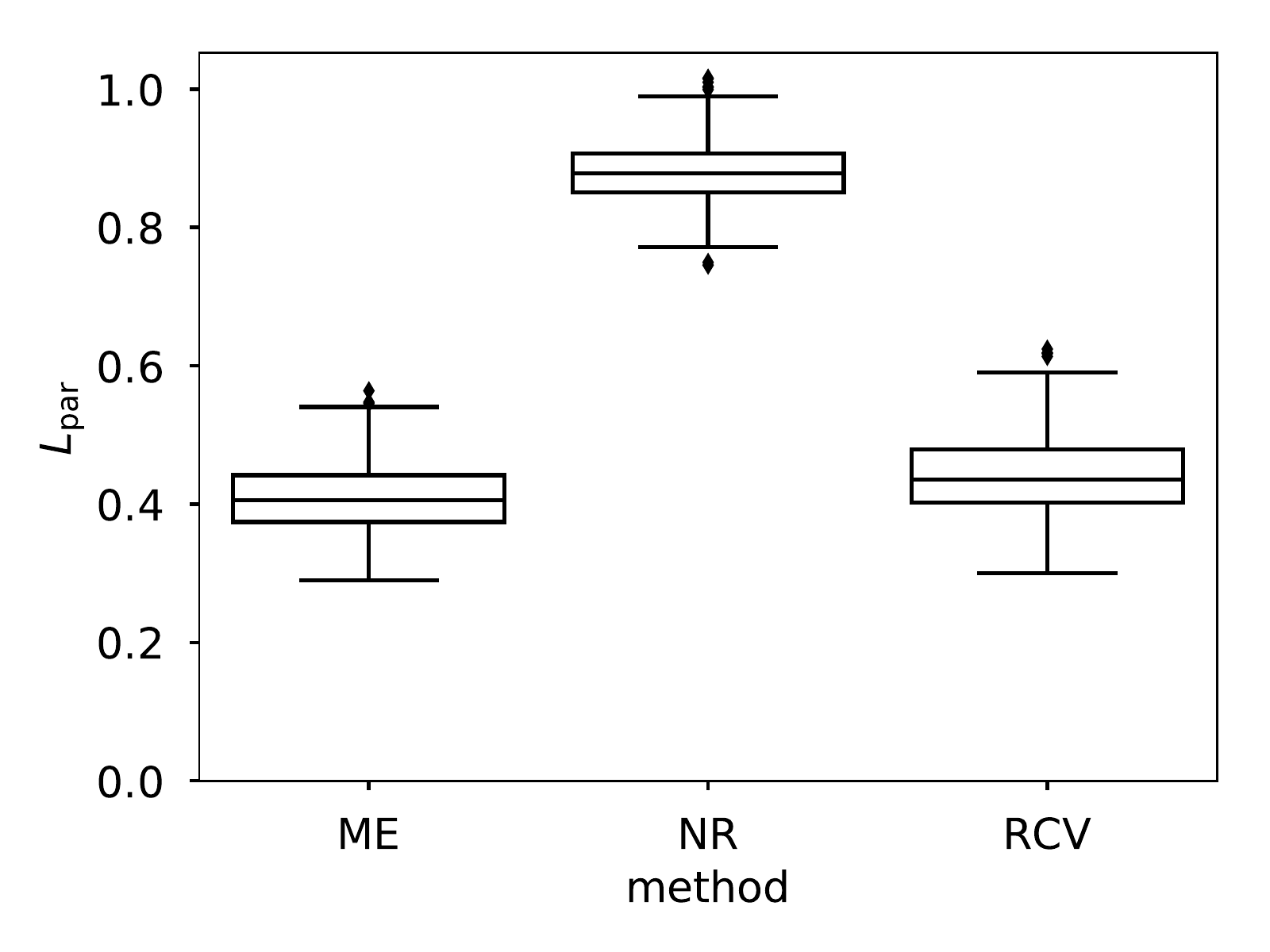}
  \end{center}
  \subcaption{$n=100$}
  \label{fig:one}
 \end{minipage}
  \begin{minipage}{0.32\hsize}
  \begin{center}
   \includegraphics[width=50mm]{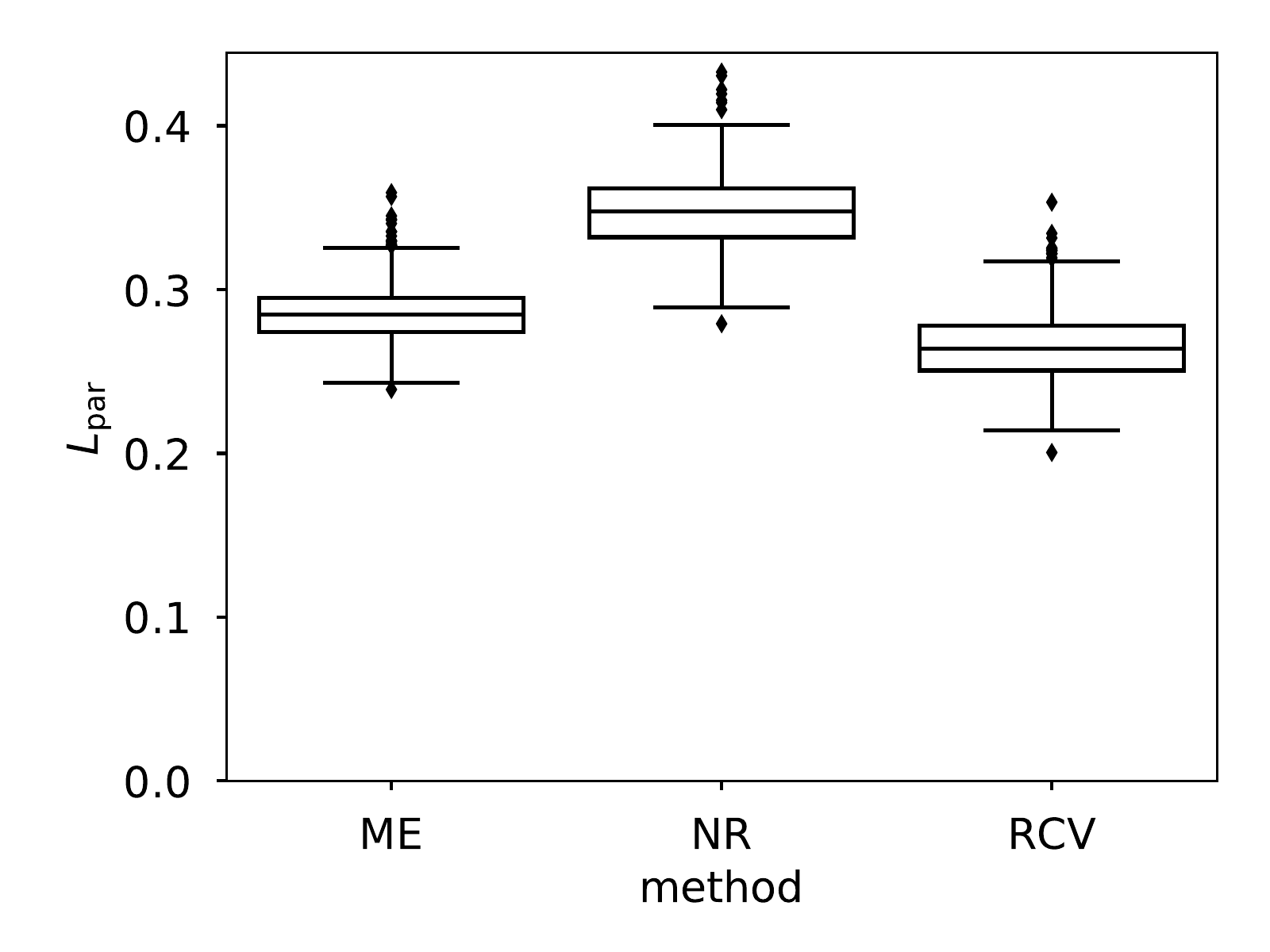}
  \end{center}
  \subcaption{$n=1000$}
  \label{fig:one}
 \end{minipage}
  \begin{minipage}{0.32\hsize}
  \begin{center}
   \includegraphics[width=50mm]{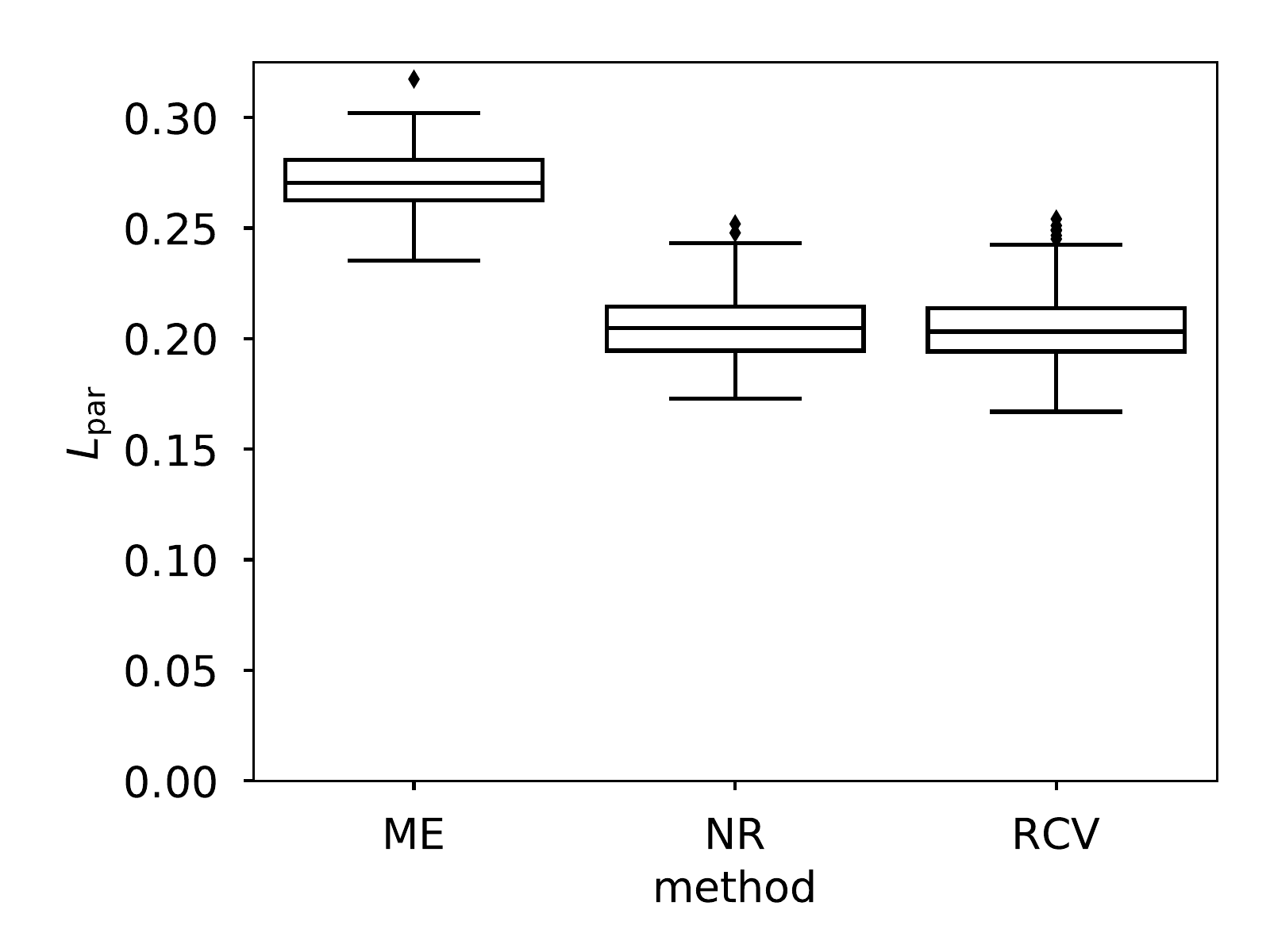}
  \end{center}
  \subcaption{$n=10000$}
  \label{fig:one}
 \end{minipage}
 \caption{Boxplots of $L_{\mathrm{par}}$ in (\ref{Lpar}) for the American Psychological Association dataset: The compared methods are the maximum entropy approach (ME), the non-regularized estimator (NR), the proposed method with cross-validation (RCV). The results with a different sample size $n$ of the train dataset are shown in (A)-(C).} 
 \label{apa}
\end{figure}

For comparison, we chose several pairs of train and test datasets randomly
to measure $L_{\mathrm{par}}$
since we do not have the true values of the model parameters nor the form of the model.
To see the dependence of the estimation performance on the sample size, 
we used different sizes ($n=100, 500, 1000, 5000, 10000$) of the train datasets,
whereas we fixed the size of the test datasets to $n=3000$.
We sampled test datasets independently $30$ times 
and sampled train datasets from the remaining data independently $30$ times 
for each size.
In calculating $L_{\mathrm{par}}$,
we used the empirical distribution of the employed test dataset as the true distribution.
For a complete ranking model,
we made the use of the likelihood of the Mallows mixture model with the number of clusters set to 3 as in \cite{busse2007cluster}.
Since R1 performs poorly in terms of $L_{\mathrm{par}}$ according to the simulation study,
we eliminate R1 from the candidate of two-fold cross-validation.

Figures \ref{apa2} and \ref{apa}
show the result.
When the sample size is small ($n=100, 500$),
the proposed method is comparable to ME,
and
NR works poorly.
When the sample size is moderate ($n=1000$),
the proposed method outperforms 
both ME and NR.
When the sample size is large ($n=5000, 10000$),
the proposed method outperforms ME,
and
it is comparable to NR.
These results indicate that
considering non-ignorable missing mechanisms
contributes to the improvement of the performance 
when the sample size is sufficient,
while the graph regularization reduces over-fitting 
when the sample size is insufficient.

\section{Conclusion}
We proposed a regularization method for partially ranked data to prevent modeling bias due to the MAR assumption and avoid over-fitting due to the complexity of missing models.
Our simulation experiments showed that the proposed method improves on the maximum entropy approach (\cite{busse2007cluster})
under non-ignorable missing mechanisms.
They also showed that the proposed method improves on the non-regularized estimator especially in estimating distribution of a partial ranking.
Our real data analysis suggested that 
moderate or large sample sizes attribute the improvement by the proposed method
and
the proposed method is effective in reducing over-fitting.

We propose two main tasks for future work.
The first task is to improve the computational efficiency of our method since it was not a priority in this study.
Leveraging partial completion of items (instead of full completion)
might be effective for reducing the computational cost.
For this purpose, the distance of top-$t$ ranking described in \cite{busse2007cluster} might be beneficial for the construction of the graph.
The second task is to develop cross-validation or an information criterion
for inferring the distribution of a latent complete ranking.
In this study,
we employed cross-validation based on the distribution of a partial ranking.
When the distribution of a latent complete ranking is of interest, cross-validation based on the distribution of a latent complete ranking would be more suitable.
However, the construction of such cross-validation would be difficult because the empirical distribution of a latent complete ranking cannot be obtained directly,
which rises ubiquitously where one uses the EM algorithm for the estimation of latent variables.
There have been several derivations of information criteria comprising the distribution of latent variables (\cite{shimodaira1994new,cavanaugh1998akaike}).
We conjecture that these derivations would be useful for inferring partially ranked data.

\appendix
\section{Algorithms}
\label{section: algorithm}
In this appendix, we provide a concise algorithm to conduct ADMM in (\ref{optvertex})-(\ref{optu}).
In the algorithm,
$\lambda$ is the regularization parameter,
$\rho$ is the penalty constant,
and
$\epsilon_p,\epsilon_d$ are two parameters for stopping the algorithm.

\begin{algorithm}[H]
\caption{Regularized estimation of $\phi$}
\label{alg:admm}
\SetKwInput{KwInput}{Input}
\SetKwInput{KwOutput}{Output}
\KwInput{$\{q_{\pi,t}^{m+1}\}_{\pi,t}$, $\phi^{0}$: the quantities (\ref{qpit}) and an initial estimate of $\phi$
}
\KwOutput{$\hat{\phi}$ : regularized estimate of $\phi$}
\SetKwFunction{Fmain}{Calc$\phi$}
\SetKwProg{Fn}{Function}{}{end}
\Fn{\Fmain{$\{q_{\pi, t}^{m+1}\}_{\pi,t}$, $\phi^0$; $\lambda$, $\rho$, $\epsilon_p, \epsilon_d$}}
{
 $(\varphi_{\pi, \pi'}^0, \varphi_{\pi',\pi}^0 )\leftarrow (\phi^0_{\pi},\phi^0_{\pi'})$ for $\{\pi,\pi'\}\in E$;\ 
 $u_{\pi, \pi'}^0, u_{\pi',\pi}^0\leftarrow 0\in\mathbb{R}^{r-1}$ for $\{\pi,\pi'\}\in E$\;
 $l\leftarrow0$; $\mathrm{res}_p\leftarrow \infty$; $\mathrm{res}_q\leftarrow \infty$;\ 
 $q_{\pi,t}\leftarrow q_{\pi,t}^{m+1}$ for $\pi\in S_r, t\in\{1,\ldots, r-1\}$\;
 \While{$\mathrm{res}_p\geq \epsilon_p$ or $\mathrm{res}_q\geq \epsilon_q$}{
 \tcc{Substitute $\argmax_{\phi}L_{\rho}(\phi,\varphi^l,u^l;q^{m+1}_{(n)})$ into $\phi^{l+1}$.}
 \ForAll {$\pi \in V$} {
 $y_{t}\gets\rho\sum_{\pi':\{\pi,\pi'\}\in E}(u_{\pi, \pi', t}^{l}-\varphi_{\pi, \pi', t}^{l})$ for $t=1,\ldots,r-1$\;
 $\nu \gets \text{Numerical solution for $\nu'$ of the equation:} \newline
 \sum_{t=1}^{r-1}\big{\{}\sqrt{(y_{ t}+\nu')^2+4\rho (r-1)q_{\pi, t}}-(y_{ t}+\nu')\big{\}}/ \{2\rho(r-1)\} =1$\;
 $\phi_{\pi, t}^{l+1}\leftarrow\big{\{}\sqrt{(y_{t}+\nu)^2+4\rho (r-1)q_{\pi, t}}-(y_{t}+\nu)\big{\}}/\{2\rho (r-1)\}$ for $t=1,\ldots,r-1$\;
 }
 \tcc{Substitute $\argmax_{\varphi}L_{\rho}(\phi^{l+1},\varphi,u^l;q^{m+1}_{(n)})$ into $\varphi^{l+1}$.}
  \ForAll {$\{\pi,\pi'\} \in E$}{
  $\phi_{\pi,\pi'}^{\ast}\leftarrow \phi_{\pi}^{l+1}+u_{\pi, \pi'}^{l}$; 
  $\phi_{\pi',\pi}^{\ast}\leftarrow \phi_{\pi'}^{l+1}+u_{\pi', \pi}^{l}$\;
  $\alpha\leftarrow[1+ \rho\|\phi_{\pi,\pi'}^{\ast}-\phi_{\pi',\pi}^{\ast}\|_2 / \{(4\lambda+\rho)\|\phi_{\pi,\pi'}^{\ast}-\phi_{\pi',\pi}^{\ast}\|_2\} ]/2$\;
$\varphi_{\pi, \pi'}^{l+1}\leftarrow\alpha\phi_{\pi,\pi'}^{\ast}+(1-\alpha)\phi_{\pi',\pi}^{\ast}$;\  
$\varphi_{\pi',\pi}^{l+1}\leftarrow\alpha\phi_{\pi',\pi}^{\ast}+(1-\alpha)\phi_{\pi,\pi'}^{\ast}$\;
}
\ForAll {$\{\pi,\pi'\}\in E$}{
$u_{\pi, \pi'}^{l+1}\leftarrow u_{\pi, \pi'}^l+(\phi_{\pi}^{l+1}-\varphi_{\pi, \pi'}^{l+1})$;
$u_{\pi',\pi}^{l+1}\leftarrow u_{\pi',\pi}^l+(\phi_{\pi'}^{l+1}-\varphi_{\pi',\pi}^{l+1})$\;
 }
 $\res_p \gets \sqrt{\sum_{\pi\in V}\sum_{\pi':\{\pi,\pi'\}\in E}\|\phi_{\pi}^{l+1} -\varphi_{\pi, \pi'}^{l+1}\|_2^2}$ ;\  
 $\res_d\gets\|\varphi^{l+1}-\varphi^{l}\|_2$\;
 $l\leftarrow l+1$\;
 }
 \Return $\phi^{l}$\;
 }
\end{algorithm}

\section{proof of proposition \ref{prop: convergence_admm}}
\label{section: proof}
\textit{Proof of Proposition \ref{prop: convergence_admm}}:
First, we express optimization (\ref{optphidecomp}) at the $m$-th step of iteration (\ref{optphi}) using an extended-real-valued function as follows:
\begin{eqnarray*}
\phi,\varphi=&\argmin_{\phi,\varphi}&f(\phi)+g(\varphi)\\
&\mathrm{s.t.} &\phi'_{\pi}=\varphi'_{\pi, \pi'} \ (\forall \{\pi,\pi'\}\in E),\
\end{eqnarray*}
where
    the functions $f:\mathbb{R}^{r!(r-1)}\rightarrow \mathbb{R}\cup \{\infty\}$ and $g:\mathbb{R}^{r!(r-1)^2}\rightarrow \mathbb{R}$ are defined as
 \begin{align*}
    f(\phi) &= \sum_{\pi\in S_r}\sum_{t=1}^{r-1}f_{\pi,t}(\phi),\\
    f_{\pi,t}(\phi)&=\begin{cases}
    \infty & (\phi_{\pi,t}\not\in [0,1] \ \text{or}\  (\phi_{\pi,t}=0 \ \text{and}\  q^{m+1}_{\pi,t}\neq 0)),\\
    0 & (\phi_{\pi,t}\in [0,1] \ \text{and}\  q_{\pi,t}^{m+1}=0),\\
    -q^{m+1}_{\pi, t}\log \phi_{\pi, t} & (\text{otherwise}),
    \end{cases}\\
    g(\varphi)&=\lambda\sum_{\{\pi,\pi'\}\in E}\|\varphi_{\pi, \pi'}-\varphi_{\pi',\pi}\|^2_2.
\end{align*}
Note that the effective domain $\mathrm{dom}(f)=\{\phi\in\mathbb{R}^{r!(r-1)}\mid f(\phi)<\infty\}$ is identical to the parameter space $\Phi=\left\{\phi \in \mathbb{R}^{r!(r-1)}: \sum_{t=1}^{r-1}\phi_{\pi, t} = 1, \phi_{\pi,t}\geq 0, \pi \in S_{r}\right\}$.
It suffices to show the following two convergences for the sequence $\{\phi^l,\varphi^l,u^l:l=0,1,\ldots \}$ generated by iteration (\ref{optvertex})-(\ref{optu}).
\begin{itemize}
    \item Residual convergence:
    the primal residual $\bar{r}^l\in\mathbb{R}^{r!(r-1)^2}$ defined by $\bar{r}_{\pi,\pi',t}^l:= \phi_{\pi,t}^l-\varphi_{\pi,\pi',t}^l$ converges to $0$ with respect to $l$:
    $\lim_{l\rightarrow\infty}\bar{r}^l=0;$
    \item Objective convergence: the convergence \[\lim_{l\rightarrow\infty}\{f(\phi^l)+g(\varphi^l)\}=\min_{\substack{\phi'\in\mathbb{R}^{r!(r-1)},\varphi'\in\mathbb{R}^{r!(r-1)^2}\\\phi_{\pi}=\varphi_{\pi,\pi'},\ (\pi,\pi'\in E)}}\{f(\phi')+g(\varphi')\}\]
    holds.
\end{itemize}

Objective convergence together with residual convergence implies convergence of the objective function $L_{\lambda}(\phi;\tau_{(n)},q^{m+1}_{(n)})$, because we have
\begin{align*}
&\left|L_{\lambda}(\phi^l;\tau_{(n)},q^{m+1}_{(n)})-\min_{\phi'\in\Phi}L_{\lambda}(\phi';\tau_{(n)},q^{m+1}_{(n)})\right|\\
&\leq |f(\phi^l)+g(\phi^l)-f(\phi^l)+g(\varphi^l)|+\left|f(\phi^l)+g(\varphi^l)-\min_{\phi'\in\Phi}L_{\lambda}(\phi';\tau_{(n)},q^{m+1}_{(n)})\right|\\
&= |g(\phi^l)-g(\varphi^l)|+\left|f(\phi^l)+g(\varphi^l)-\min_{\phi'_{\pi}=\varphi'_{\pi,\pi'},\ (\pi,\pi'\in E)}\{f(\phi')+g(\varphi')\}\right|\\
&\rightarrow 0\ (l\rightarrow \infty),
\end{align*}
where it follows 
from residual convergence and the continuity of $g$
that $|g(\phi^l)-g(\varphi^l)|\rightarrow 0$. 

The following is a sufficient condition for objective and residual convergence based on ADMM (see Section 3.2 of \cite{boyd2011distributed}):
\begin{enumerate}[(I)]
    \item The functions $f$ and $g$ are closed, proper, and convex; \label{big1}
    \item Unaugmented Lagrangian $\tilde{L}_0$ has a saddle point. \label{big2}
\end{enumerate}
Here unaugmented Lagrangian $\tilde{L}_0$ is 
defined as
\begin{align*}
\tilde{L}_{0}(\phi,\varphi,y;q^{m+1}_{(n)})=f(\phi)+g(\varphi)+ \sum_{\pi\in V}\sum_{\{\pi,\pi'\}\in E}y_{\pi,\pi}^{\mathrm{T}}(\phi_{\pi}-\varphi_{\pi,\pi'}).
\end{align*}
In what follows, we show that conditions (\ref{big1}) and (\ref{big2}) hold.

\textit{Confirming condition (\ref{big1}):} $g$ is clearly a closed, proper, and convex function because $g$ is a positive quadratic function.
Each function $f_{\pi,t}\ (\pi\in S_r, t\in \{1,\ldots,r-1\})$ is closed because every level set $V_{\gamma}=\{x\in \mathbb{R}^{r!(r-1)}\mid f_{\pi,t}(\phi)\}$ with $\gamma\in\mathbb{R}$ is a closed set:
\[
V_{\gamma}=\begin{cases}
(-\infty,\infty)\times\ldots \times[\exp(-\gamma),1]\times\ldots\times(-\infty,\infty) & (\gamma\geq 0\ \text{and}\ q_{\pi,t}\neq 0),\\
(-\infty,\infty)\times\ldots \times[0,1]\times\ldots\times(-\infty,\infty) & (\gamma\geq 0\ \text{and}\ q_{\pi,t}= 0),\\
\emptyset & (\text{otherwise}).
\end{cases}
\]
Therefore, $f$ is closed.
$f$ is proper because $f\geq 0>-\infty$ everywhere and $f(\phi)<\infty$ for $\phi\in\mathbb{R}^{r!(r-1)}$ satisfying $\phi_{\pi,t}=1/(r-1),\ \pi\in S_r,t=1,\ldots,r-1$.
Each function $f_{\pi,t}$ $(\pi\in S_{r}, t\in \{1,\ldots,r-1\})$ is convex because the effective domain $\mathrm{dom}(f_{\pi,t})$ is a convex set and $\nabla^2f_{\pi,t}(\phi)$ is positive semidefinite for all $\phi\in\mathrm{dom}(f_{\pi,t})$.
Therefore, $f$ is convex.
Thus, condition (\ref{big1}) holds.

\textit{Confirming condition (\ref{big2}):} We employ the following sufficient condition for the existence of a saddle point described as Assumption 5.5.1 and Proposition 5.5.6 in Section 5.5 of \cite{bertsekas2015convex}:
\begin{enumerate}[(i)]
    \item For each $\phi\in\mathbb{R}^{r!(r-1)},\varphi\in\mathbb{R}^{r!(r-1)^2}$, $-\tilde{L}_0(\phi,\varphi,\cdot): \mathbb{R}^{r!(r-1)^2}\rightarrow \mathbb{R}\cup\{\infty\}$ is convex and closed; \label{small1}
    \item For each $y\in\mathbb{R}^{r!(r-1)^2}$, $\tilde{L}_0(\cdot,\cdot,y): \mathbb{R}^{r!(r-1)}\times\mathbb{R}^{r!(r-1)^2}\rightarrow \mathbb{R}\cup\{\infty\}$ is convex and closed; \label{small2}
    \item Functions $L^+$ and $L^-$ are proper,
    where $L^{+}:\mathbb{R}^{r!(r-1)}\times\mathbb{R}^{r!(r-1)^2}\rightarrow \mathbb{R}\cup\{\infty\}$ and $L^{-}:\mathbb{R}^{r!(r-1)^2}\rightarrow \mathbb{R}\cup\{\infty\}$ are defined as
\begin{align*}
    L^{+}(\phi,\varphi)=\sup_{y\in \mathbb{R}^{r!(r-1)^2}}\tilde{L}_0(\phi,\varphi,y)
    \text{ and }
    L^{-}(y) = \sup_{\substack{\phi\in\mathbb{R}^{r!(r-1)}\\\varphi\in\mathbb{R}^{r!(r-1)^2}}}-\tilde{L}_0(\phi,\varphi,y);
\end{align*}
    \label{small3}
    \item For each $\gamma\in\mathbb{R}$, the level set $\{\phi,\varphi\mid L^+(\phi,\varphi)\leq \gamma\}$ is compact; \label{small4}
    \item For each $\gamma\in\mathbb{R}$, the level set $\{y\mid L^-(y)\leq \gamma\}$ is compact. \label{small5}
\end{enumerate}

Condition (\ref{small1}) holds because $-\tilde{L}_0(\phi,\varphi,\cdot)$ is linear for $\phi\in\mathrm{dom}(f)$) and $-\infty$ for $\phi\not\in\mathrm{dom}(f)$. 
Condition (\ref{small2}) holds because $\tilde{L}_0(\cdot,\cdot,y)$ is the sum of convex and closed functions.

To confirm condition (\ref{small3}), we will show that $L^+$ and $L^-$ are proper.
Set $\phi^{\ast}\in \mathbb{R}^{r!(r-1)}$ and $\varphi^{\ast}\in\mathbb{R}^{r!(r-1)^2}$ such that
\begin{align*}
\phi^{\ast}_{\pi,t} &= 1/(r-1) \text{ for all }\pi\in V, t\in\{1,\ldots,r-1\} \text{ and }\\
\varphi^{\ast}_{\pi,\pi',t} &= 1/(r-1)\text{ for all }\{\pi,\pi'\}\in E, t\in\{1,\ldots,r-1\}.
\end{align*}
It follows that $L^+$ is proper since 
\begin{align*}
L^+(\phi,\varphi) &\geq \tilde{L}_0(\phi,\varphi,0)\geq 0>-\infty \text{ for all } \phi,\varphi
\text{ and}\\
L^+(\phi^{\ast},\varphi^{\ast})&=\tilde{L}_0(\phi^{\ast},\varphi^{\ast},0)<\infty \text{ for } (\phi,\varphi) = (\phi^{\ast},\varphi^{\ast}).
\end{align*}
It follows that $L^-$ is proper since 
\begin{align*}
L^-(y)&\geq -\tilde{L}_0(\phi^{\ast},\varphi^{\ast},y)>-\infty\text{ for all }y \text{ and }\\
L^-(0) &= \sup_{\phi,\varphi}\{-f(\phi)-g(\varphi)\}\leq 0<\infty \text{ for }y=0.
\end{align*}
Therefore, condition (\ref{small3}) is confirmed.

To confirm conditions (\ref{small4}) and (\ref{small5}), it suffices to show that the level sets are closed and bounded.
Since the function obtained by taking the point-wise supremum of a family of closed functions is again closed, both $L^+$ and $L^-$ are closed and thus their level sets are closed. The remaining part of the proof is to show that the level sets of $L^{+}$ and  $L^{-}$ are bounded.

We will show that all level sets of $L^+$ are bounded by focusing on the effective domain of $L^+$.
We show that the effective domain $\mathrm{dom}(L^+)$ is a subset of the bounded set 
\[
B = \left\{\phi\in\mathbb{R}^{r!(r-1)},\varphi\in\mathbb{R}^{r!(r-1)^2}\mid \phi_{\pi,t}\geq 0, \sum_{t=1}^{r-1}\phi_{\pi,t}=1, \phi_{\pi,t}-\varphi_{\pi,\pi',t}=0,\ \{\pi,\pi'\}\in E\right\},
\]
according to which all level sets of $L^{+}$ are bounded.
If $\phi_{\tilde\pi,\tilde{t}}-\varphi_{\tilde\pi,\tilde\pi',\tilde{t}}\neq 0$ for some $\{\tilde\pi,\tilde\pi'\}\in E$ and $\tilde{t}\in\{1,\ldots,r-1\}$, we can take a sequence $\{y^n\}_{n=1}^{\infty}\subset \mathbb{R}^{r!(r-1)^2}$ such that $y^n_{\tilde\pi,\tilde\pi',\tilde{t}}=n(\phi_{\tilde\pi,\tilde{t}}-\varphi_{\tilde\pi,\tilde\pi',\tilde{t}})$ for $(\{\pi,\pi'\},t)= (\{\tilde\pi,\tilde\pi'\},\tilde{t})$ and $y^n_{\pi,\pi',t}=0$ otherwise.
For the sequences $\{y^n\}$, we have 
\[\lim_{n\rightarrow\infty}\sum_{\pi\in V}\sum_{\{\pi,\pi'\}\in E}(y_{\pi,\pi'}^{n})^{\mathrm{T}}(\phi_{\pi}-\varphi_{\pi,\pi'})=\lim_{n\rightarrow\infty}n(\tilde\phi_{\pi,\tilde{t}}-\varphi_{\tilde\pi,\tilde\pi',\tilde{t}})^2= \infty,\]
from which we obtain $L^+(\phi,\varphi)=\sup_{y\in\mathbb{R}^{r!(r-1)^2}}\tilde{L}_0(\phi,\varphi,y)=\infty$.
Therefore, the effective domain of $L^+(\phi,\varphi)$ is included in the bounded set $B$ and thus all level sets of $L^+$ are bounded.

We will show that all level sets of $L^-$ are bounded by showing that $L^{-}$ is coercive, i.e., for any sequence $\{y^n\}_{n=1}^{\infty}\subset \mathbb{R}^{r!(r-1)^2}$ satisfying $\lim_{n\rightarrow\infty}\|y^n\|_2=\infty$,
we have
$\lim_{n\rightarrow\infty}L^-(y^n)=\infty.$
For any given sequence $\{y^n\}_n$ satisfying $\lim_{n\rightarrow\infty}\|y^{n}\|_{2}=\infty$, 
take sequences $\{\phi^n\}_n$ and $\{\varphi^n\}_n$ 
such that 
$\phi_{\pi,t}^n=1/(r-1)$  and $\varphi_{\pi,\pi'}^n = \phi_{\pi}^n+y_{\pi,\pi'}^n/\|y_{\pi,\pi'}^n\|_2.$
For sequences $\{\phi^n\}_n$ and $\{\varphi^n\}_n$, we obtain 
\begin{align*}
    &-f(\phi^n)=\sum_{\pi\in V}\sum_{t=1}^{r-1}q^{m+1}_{\pi, t}\log (r-1)\geq 0,\\
    &-g(\phi^n)=-\sum_{\pi\in V}\sum_{\{\pi,\pi'\}\in E}\left\|\frac{y_{\pi,\pi'}^n}{\|y_{\pi,\pi'}^n\|_2}-\frac{y_{\pi',\pi}^n}{\|y_{\pi',\pi}^n\|_2}\right\|^2_2\geq -2r!(r-1),\\
    &-\sum_{\pi\in V}\sum_{\{\pi,\pi'\}\in E}(y_{\pi,\pi'}^{n})^{\mathrm{T}}(\phi_{\pi}^n-\varphi_{\pi,\pi'}^n)=\sum_{\pi\in V}\sum_{\{\pi,\pi'\}\in E}\|y_{\pi,\pi'}^{n}\|_2\geq\|y^n\|_2.
\end{align*}
Hence, $L^-$ is coercive since we have $L^-(y^n)\geq -\tilde{L}_0(\phi^n,\varphi^n,y^n)\geq -2r!(r-1)+\|y^n\|_2\rightarrow \infty\ (n\rightarrow\infty)$ for any sequence $\{y^{n}\}_{n=1}^{\infty}$ satisfying $\|y^n\|_2\rightarrow\infty$, and thus all level sets of $L^-$ are bounded.

From the above, conditions (\ref{small4}) and (\ref{small5}) are satisfied and thus 
we complete the proof.
\qed

\bibliographystyle{plainnat}
\bibliography{NakamuraYanoKomaki}
\end{document}